\documentclass[fleqn,usenatbib]{mnras}

\usepackage{newtxtext,newtxmath}

\usepackage[T1]{fontenc}
\usepackage{ae,aecompl}

\usepackage{graphicx}	
\usepackage{amsmath}	
\usepackage{amssymb}	
\usepackage{threeparttable}
\usepackage{pdflscape}
\usepackage{longtable}
\usepackage{booktabs}
\usepackage{multirow}
\usepackage{caption}
\usepackage{subcaption}
\usepackage{arydshln}
\usepackage{times}
\usepackage[export]{adjustbox}
\usepackage[usenames,dvipsnames]{color}
\definecolor{gray}{gray}{0.5}
\usepackage[bottom]{footmisc}
\newcommand{\kms}{km s$^{-1}$}
\newcommand{\lxlbol}{log$L_{\rm X}/L_{\rm bol}$}
\newcommand{\excl}{BPMG$_{\rm Excl}$}
\newcommand{\incl}{BPMG$_{\rm Incl}$}
\newcommand{\unifexcl}{BPMG$_{\rm Unif,Excl}$}
\newcommand{\unifincl}{BPMG$_{\rm Unif,Incl}$}

\newcommand{\listone}{{\it BANYAN II list}}
\newcommand{\listexc}{{\it exclusive list}}
\newcommand{\listinc}{{\it inclusive list}}
\newcommand{\listexcinc}{{\it exclusive \rm and \it inclusive lists}}
\newcommand{\listthree}{{\it SIMBAD list}}
\newcommand{\casei}{{\it Case I}}
\newcommand{\caseii}{{\it Case II}}
\newcommand{\caseiiex}{{\it Case II (Excl.)}}
\newcommand{\caseiiin}{{\it Case II (Incl.)}}
\newcommand{\caseiiiex}{{\it Case III (Excl.)}}
\newcommand{\caseiiiin}{{\it Case III (Incl.)}}
\newcommand{\caseiii}{{\it Case III}}
\newcommand{\caseiv}{{\it Case IV}}
\newcommand{\casev}{{\it Case V}}

\title[Bayesian moving group memberships]{Bayesian assessment of moving group membership: importance of models and prior knowledge}

\author[J. Lee and I. Song]{
Jinhee Lee\thanks{E-mail: jinhee@uga.edu}
and Inseok Song\thanks{E-mail: song@uga.edu}
\\
Department of Physics and Astronomy, The University of Georgia, Athens, GA 30602-2451\\
}

\date{Accepted XXX. Received YYY; in original form ZZZ}

\pubyear{2017}

\begin{document}
\label{firstpage}
\pagerange{\pageref{firstpage}--\pageref{lastpage}}
\maketitle

\begin{abstract}
Young nearby moving groups are important and useful in many fields of astronomy such as studying exoplanets, low-mass stars, and the stellar evolution of the early planetary systems over tens of millions of years, which has led to intensive searches for their members.
Identification of members depends on the used models sensitively, 
 therefore, careful examination of the models is required.
 In this study, we investigate the effects of the models used in moving group membership calculations based on a Bayesian framework (e.g., BANYAN II) focusing on the beta-Pictoris moving group (BPMG).
Three improvements for building models are suggested:
 (1) updating a list of accepted members by re-assessing memberships in terms of position, motion, and age, 
 (2) investigating member distribution functions in $XYZ$, and
 (3) exploring field star distribution functions in $XYZ$ and $UVW$. 
 The effect of each change is investigated, and we suggest using all of these improvements simultaneously in future membership probability calculations.
 Using this improved MG membership calculation and the careful examination of the age, 57 bona fide members of BPMG are confirmed including 12 new members.
 We additionally suggest 17 highly probable members.

\end{abstract}

\begin{keywords}
methods: statistical -- stars: kinematics and dynamics -- open clusters and associations: general -- open clusters and associations: individual: $\beta$-Pictoris moving group
\end{keywords}



\section{Introduction}

Young nearby moving groups (MGs hereafter) are sparse stellar associations whose members were formed together in loose environment and share common proper motions.
Therefore, members of MGs spread out over time in space, and, after hundreds of millions of years, they are not easily distinguishable against old field stars.
After the first identification of the nearby young MG, TW Hydrae Association \citep{kas97}, about ten additional MGs were identified \citep{web99, zuc00, tor00, zuc01a, zuc04, tor08, zuc11}.
 
Nearby, young MGs are unique objects in astronomy because of their proximity and youth.
MG members are prime targets for exoplanet imaging \citep{lag10, cha04, mar08}, because orbiting young planets are brighter and more widely separated compared to those around old, distant stars.
They are also useful in calibrating stellar age-dating methods and in studying the evolution of low-mass stars (Binks \& Jeffries 2016; Malo et al. 2014; Murphy, Lawson \& Bento 2015).
In addition, they are essential to understanding the recent star formation history in the solar neighbourhood \citep{schn12, tor03}.
These studies are critically dependant on assignments of MG membership because they rely on MG group properties such as age, space motion, and location from Earth that are derived from known MG members.

Because of their importance, identification of MG members has been intensively investigated (Song, Zuckerman \& Bessell 2003; Zuckerman \& Song 2004; Torres et al. 2006; Schlieder, L\'{e}pine \& Simon 2010; Kiss et al. 2011; Rodriguez et al. 2011; Shkolnik et al. 2012; Malo et al. 2013; Gagn\'{e} et al. 2014; Murphy et al. 2015). 
Among methods applied to searching for MG members, a statistical approach based on a Bayesian framework has become popular recently \citep{mal13, gag14}.
While there are many advantages in using the Bayesian method (e.g., an intuitively easier interpretation of the result due to the quantitative membership probability and the availability of marginalising over nuisance parameters), the resultant membership probability needs to be carefully adopted because of its sensitive dependence on input models.  

In this paper, we examine the impact of models and prior knowledge in the Bayesian MG membership probability, focusing on the accepted member list and the distribution functions for Beta-Pictoris moving group (BPMG hereafter) and field stars.
We then suggest improvements to the membership calculations, and finally provide a list of confirmed and probable members of the BPMG.

\section{Bayesian moving group membership calculations}
Our method is based on the same Bayesian principle as used in BANYAN II \citep{gag14}, and one of our main purposes of this paper is to demonstrate the careful treatments of model parameters.
Therefore, to minimize any possible confusion, we decide to use the BANYAN II notation in describing various terms related to the Bayesian MG membership probability calculation.
Throughout this paper, in developing, validating, and comparing our method and result, we used BPMG as the main test case because this MG is one of the youngest, nearest MGs with many discovered members spread over a large area of the sky.
For various purposes, different lists of stars for BPMG were used (see Table~\ref{tab1}).

\subsection{Validation of our calculation: comparison against the BANYAN II result}
\begin{figure}
\includegraphics[width=0.9\columnwidth]{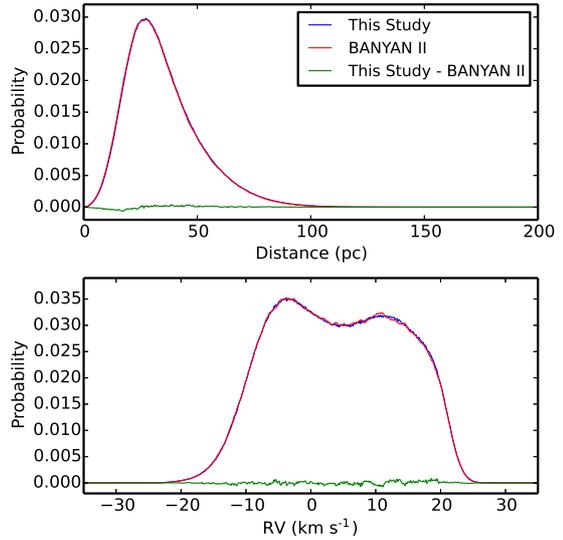}
\caption{Prior probability distribution functions (PDFs) of RV and distance for BPMG from BANYAN II (red) and those from this study (blue) assuming identical model parameters.
The differences of two prior PDFs are presented with green lines.}
\label{fig1}
\end{figure}

We developed a \textsc{python} script to calculate Bayesian MG membership probability.
Bayesian probability (the posterior probability) is proportional to the product of the prior probability and the likelihood, which  
are both derived from models for MGs and field stars.
To validate our code, we compare our membership calculation results (the Bayesian probability) against those of BANYAN II using identical parameters.
Prior probability distribution functions (PDFs) for observables such as distance, radial velocity (RV), magnitude of proper motion, and galactic latitude are reproduced, and two of these are compared in Fig.~\ref{fig1}.
Prior PDFs from BANYAN II and those from this study are similar but not exactly the same because these prior PDFs are generated from random simulated stellar distributions.
We also used the BANYAN II group size ratios between MGs and the field stars.

\begin{figure}
\includegraphics[width=0.9\columnwidth]{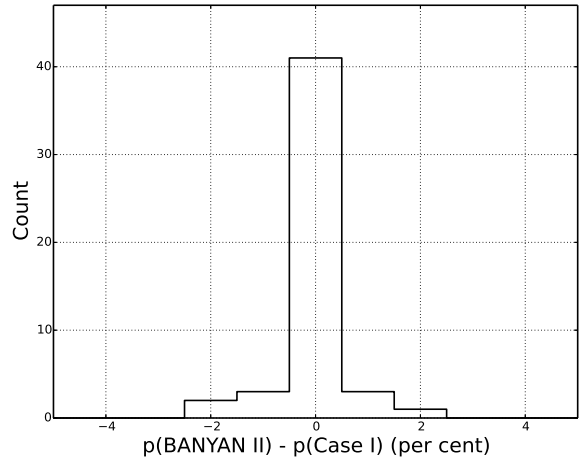}
\caption{A comparison of membership probabilities from BANYAN II and those from this study (\casei) utilizing BPMG members from the \listone\ (Table~\ref{tab1}).
Both calculations are based on the same models (Table~\ref{tab2}).}
\label{fig2}
\end{figure}

The membership probability calculations from BANYAN II and those from our code are compared in Fig.~\ref{fig2} using a list of BPMG members taken from the BANYAN II webpage (\listone, see Table~\ref{tab1}).
This figure shows that our code replicates almost the same result of BANYAN II except for a handful of abnormal cases where the difference between the two calculations is likely caused by the small difference in prior PDFs shown in Fig.~\ref{fig1}.

\begin{landscape}
\begin{table}
\caption{Star lists used in this study.}
\label{tab1}
\begin{threeparttable}
\begin{tabular}{p{.1\linewidth}p{.035\linewidth}p{.35\linewidth}p{.445\linewidth}}
\hline
Name & N & Description & Purpose of usage \\ \hline
\listone  & 50 & A list of previously known BPMG members taken from the BANYAN II webpage \newline (www.astro.umontreal.ca/$\sim$gagne/banyanII.php; \citet{gag14}). & To validate our calculation by comparing it to BANYAN II's calculation (Section 2.1). \newline To create lists of initially accepted BPMG members (\listexc\ and \listinc) (Section 2.2.1). \\
\hline
\listexc  \newline \listinc & 39 \newline 47 & Lists of initially accepted BPMG members based on the membership assessment criteria (exclusive and inclusive).   \newline Subsets of the \listone. & To construct improved BPMG models (Section 2.2.1 and 2.2.2). \\
\hline
\listthree & 275 & A list of BPMG members in SIMBAD database as of 2017 April \newline (http://simbad.u-strasbg.fr/simbad/), which constitutes the whole \listone. & To test effects of models (Section 3.1).  \newline To provide the updated list of bona fide and probable members of BPMG (Section 3.2).\\
\hline
bona fide member list & 51(57\tnote{a} ) & A confirmed BPMG member list in this study (Table~\ref{tab5}). & To derive revised model parameters for BPMG (Section 3.2). \\
\hline
\end{tabular}
\begin{tablenotes}
\item[a] Including 6 classical BPMG members showing moderate signs of youth, discussed in section 3.2.1.
\end{tablenotes}
\end{threeparttable}
\end{table}

\begin{table}
\caption{Combinations of different factors in building models.}
\label{tab2}
\begin{threeparttable}
\begin{tabular}{llcccccccc}
\hline
\multirow{2}{*}{Model factors} & \multirow{2}{*}{Type} & & \multicolumn{7}{c}{cases in this study} \\
\cline{4-10} 
& & BANYAN II & \casei\ & \caseiiex\ & \caseiiin\  & \caseiiiex\ & \caseiiiin\ & \caseiv\ & \casev\ \\
\hline
\multirow{3}{*}{Selection of BPMG members} & BANYAN II  & O & O &   &   &   &  &O&  \\ \cdashline{2-10}
 & Exclusive    &     &    & O&  &O &   &  &O\\ \cdashline{2-10}
                           & Inclusive    &      &    &   &O&   &O&   &   \\
\hline
\multirow{2}{*}{$XYZ$ distribution function of BPMG members}   & Gaussian    & O  &O &O&O&   &   &O& \\ \cdashline{2-10}
  & Uniform      &      &    &   &   &O&O&   &O\\
\hline
\multirow{2}{*}{Distribution models of field stars} & BANYAN II & O & O&O&O&O&O&   &\\ \cdashline{2-10}
                           & A new model             &     &    &  &    &  &   &O&O\\
\hline
\end{tabular}
\end{threeparttable}
\end{table}
\end{landscape}

\subsection{Improvement over BANYAN II: updating models}

Different models modify likelihoods and prior PDFs, making changes in membership probabilities. 
In this study, we consider three important factors in building models:  
(1) a list of adopted initial MG members, 
(2) a distribution function of BPMG members in $XYZ$, and 
(3) new distribution functions of field stars in $XYZ$ and $UVW$ \footnote{$U$, $V$, and $W$ are positive toward the directions of the Galactic centre, Galactic rotation, and the North Galactic Pole, respectively..
To investigate the effects of these three factors, we carried out various combinations of them and compared the result against that of BANYAN II (see Table~\ref{tab2}).}

\subsubsection{Re-examination of MG membership}
To define the characteristics of a MG, one has to start with a certain list of   MG members.
For example, to measure the extent of the distribution of MG members in $XYZ$ and $UVW$, one has to model the distribution of accepted members in such 3D spaces, which means that a different set of stars will produce different distribution model parameters for the MG.
This seemingly straightforward task of establishing a list of accepted MG members is more complicated because the assignment of membership status is an iterative process.
Starting with a stringent initial list of accepted members, a MG will have tighter distributions in $XYZ$ and $UVW$ which in turn forces any candidate members need to fit the tighter parameters.
On the other hand, if the membership assessment starts with a more lenient list of members, the distribution model of the MG will become more inclusive, accepting more marginal members as true members.

In this section, we reassess membership of previously known BPMG members from the \listone.
A true member should possess not only similar kinematic characteristics ($XYZ$ and $UVW$) but also a similar age with other members.

Firstly, we tried to flag outliers in $XYZ$ and $UVW$ by calculating standard deviations ($\sigma$) in each $X, Y, Z, U, V$, and $W$.
To calculate standard deviations that properly represent the MG as a whole, we excluded obvious outliers that are above the upper limit \footnote{(3rd quartile [75\%-ile] + 1.5$\times$ interquartile range (IQR))} or below the lower limit \footnote{(1st quartile [25\%-ile] - 1.5$\times$ IQR)}. 
HIP 11360 and $\eta$ Tel A deviate $\gtrsim$5$\sigma$ in $UVW$ from the mean.
Including $\eta$ Tel B, these 3 stars were determined to be kinematic outliers.
Additionally, HIP 50156 was determined as a marginal outlier because this star has Z value about 4.5$\sigma$ away from the mean.

Age outliers were determined based on age indicators.
 Genuine BPMG members should show clear signs of youth appropriate to the age of BPMG (12$-$25 Myr: \cite{zuc04, ort02, bin16, mam14}).
Young ($<$100 Myr), Sun-like or lower mass (mid-F to M types) stars can be distinguished from older counterparts because they have brighter photometric magnitude, NUV excess, X-ray excess, and/or strong Li absorption \citep{sod10, rod11, zuc04, bin16}.
Fourty-two out of 50 stars in the \listone\ were estimated to be $\lesssim$25 Myr.
Among the remaining 8 stars, no age-related data exists for $\eta$ Tel B, 
and the age estimation for the other 7 stars involves significant uncertainties (HIP 10679, HIP 10680, 2M J06085283-2753583, HIP 21547, HIP 88726A, HIP 88726B, and HIP 92024).
Both HIP 10679 and HIP 10680 show strong Li absorption features but can be $\sim$100 Myr old based on a $V-K$ versus $M_K$ colour-magnitude diagram (CMD).
The location of 2M J06085283-2753583 on the CMD is ambiguous, making it difficult to unambiguously assess if this star is $\lesssim$ 25 Myr.
Because HIP 21547, HIP 88726 A, HIP 88726 B, and HIP 92024 have early spectral types (F0 to A5), reliable age estimates are difficult.
Even though the age estimates of these 7 stars are ambiguous, none of them are marked as evident old star interlopers.

\begin{figure}
\includegraphics[width=0.99\columnwidth]{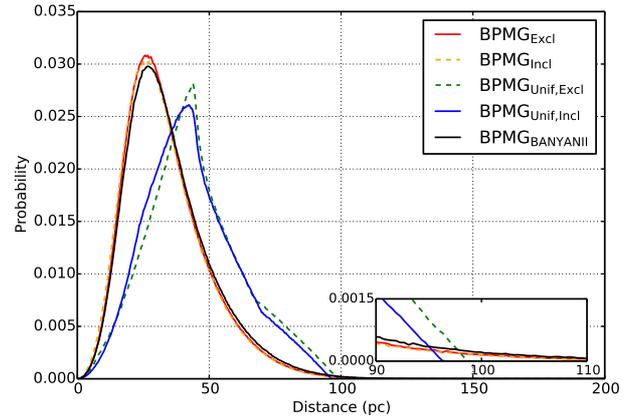}
\caption{Prior PDFs of distance for BPMG models. 
Inset is an enlarged view of probabilities at around 100 pc. 
Models using minimum volume enclosing ellipsoid (uniform distribution in $XYZ$; \unifexcl\ and \unifincl$-$\caseiii) reach zero probability at around 100 pc,  while other models, assuming Gaussian distribution, do not.}
\label{fig3}
\end{figure}

Based on age and kinematic data, one can try to build two different extreme cases for the initial accepted BPMG member list: (1) an \listinc\ containing all marginal members with large age uncertainties and outlying kinematics and (2) an \listexc\ containing only unambiguous members in age and kinematics. 
Using these two lists, parameters for two BPMG distribution models (\excl\ and \incl\ in Table~\ref{tab3}) are estimated  
 by fitting a single Gaussian model utilizing the Gaussian mixture model algorithm from \textsc{python scikit-learn} package \citep{ped11}. 
The distance prior PDFs for both models and 2D projections of \excl\ are presented in  Fig.~\ref{fig3}, ~\ref{fig5}, ~\ref{fig6}, and ~\ref{fig7}.

\begin{figure*}
  \includegraphics[width=0.99\textwidth]{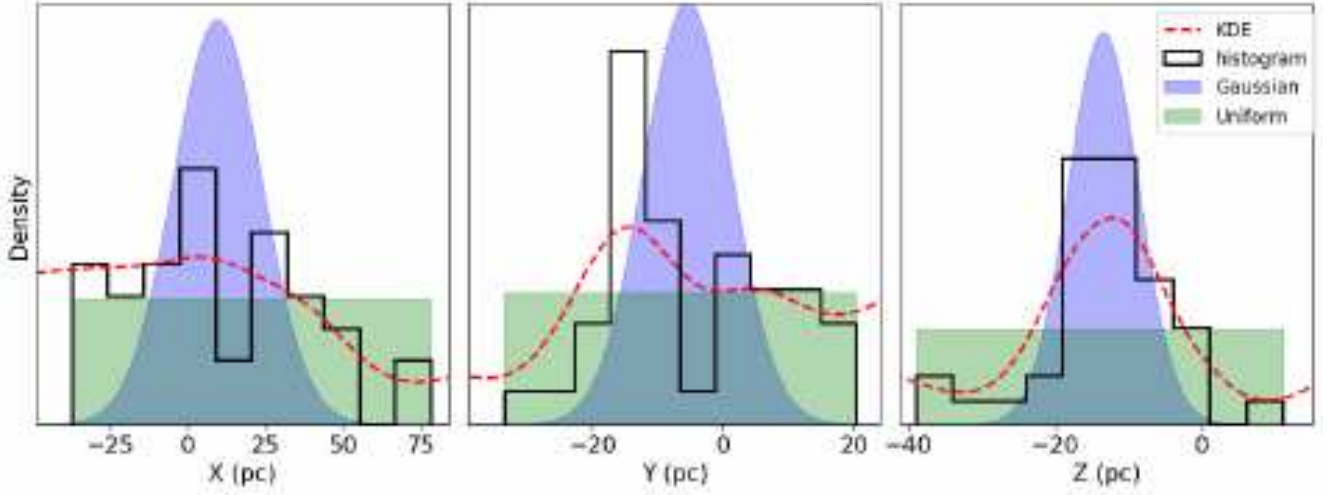}
 \caption{Normalized density of stars in the \listexc\ in each $X$, $Y$, and $Z$.
The density function using a kernel density estimation (KDE), which shows binning-free distribution, is presented with a red dashed line overlaid on the data histogram.
In the boundary region, KDE underestimates the density because there is no data (boundary effects) and this boundary effects are corrected by truncating the kernel at the outermost boundaries (minimum and maximum values for each $X$, $Y$, $Z$).
Gaussian and uniform distribution of pseudo data are presented as shaded area with blue and green colors, respectively.
}
\label{fig4}
\end{figure*}

\begin{figure*}
\begin{subfigure}{0.66\textwidth}
    \includegraphics[width=\textwidth]{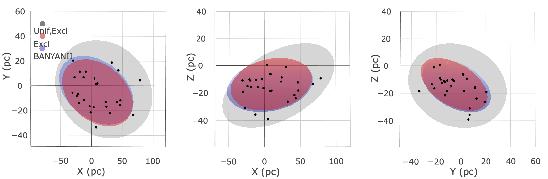}
\end{subfigure}
\begin{subfigure}{0.33\textwidth}
    \includegraphics[width=\textwidth]{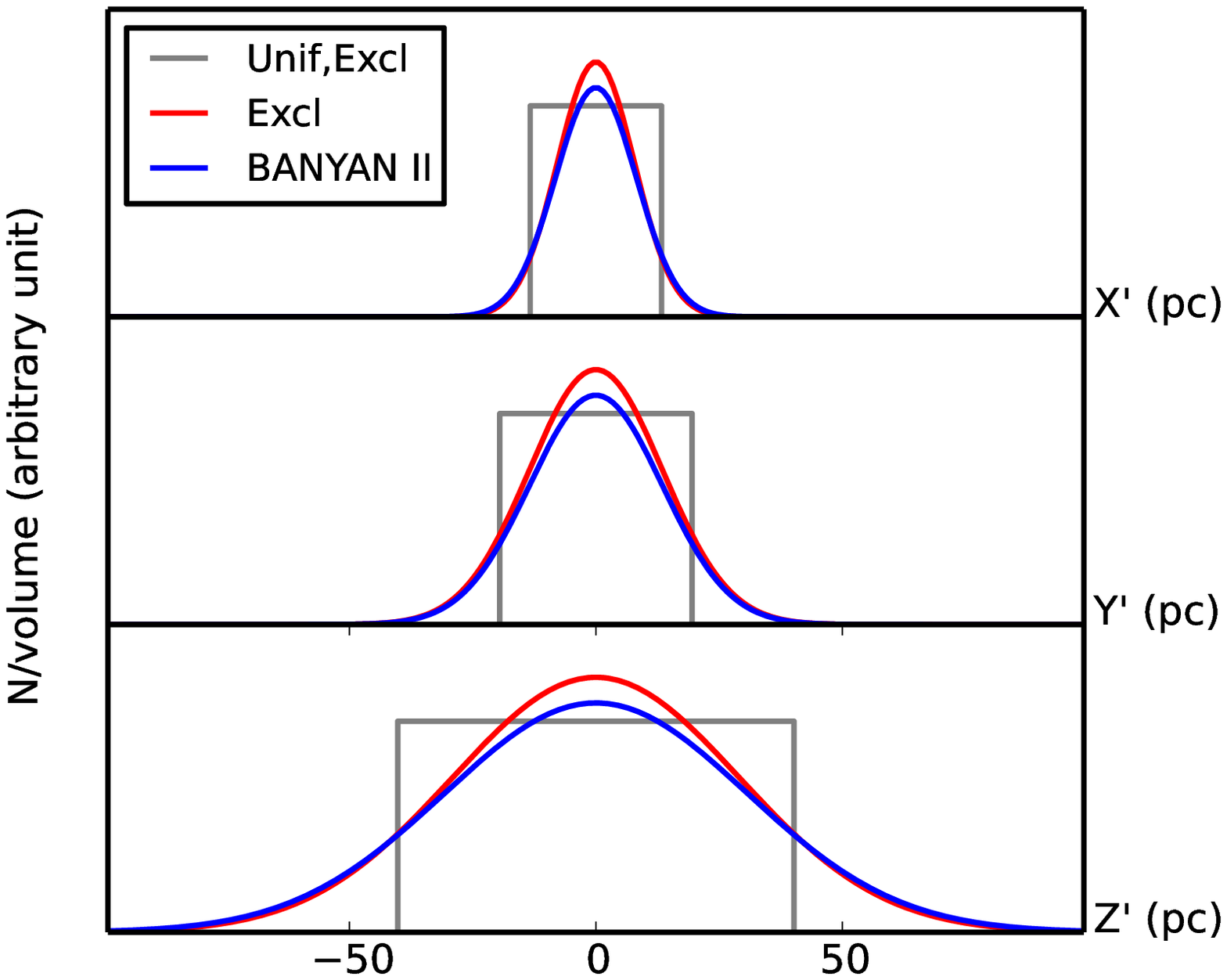}
\end{subfigure}
\caption{2D projections of BPMG models in $XYZ$.
BPMG$\rm_{BANYANII}$, \excl, and \unifexcl\ models are presented with blue, red, and grey, respectively.
Stars in the \listexc\ are presented as dots.
We plotted 1.2 times the MVEE semi-major axes for \unifexcl\ and 2-$\sigma$ boundaries for the other two Gaussian models.
The former should have 100 per cent of integrated probability within the plotted boundaries while the latter should have 87 per cent of integrated probabilities within the plotted boundaries.
Right panel presents the number density (i.e., probability density function) along principal major axes ($X'$, $Y'$, and $Z'$) for each model.}
\label{fig5}
\end{figure*}

\begin{figure*}
  \includegraphics[width=\textwidth]{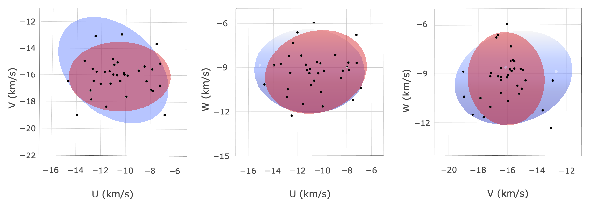}
\caption{2D projections of BPMG models in $UVW$.
BPMG$\rm_{BANYANII}$ and \excl\ models are presented with blue and red, respectively. 
Stars in the \listexc\ are presented as dots.
The model boundaries are 2-$\sigma$ values in each semi-major axis.}
\label{fig6}
\end{figure*}

\subsubsection{$XYZ$ Distribution models of BPMG members}
In this section, we discuss a proper probability distribution of MG members as a function of distance from the centre of the MG.
Should a star closer to the MG centre be assigned with a higher membership probability, or should any star within a given distance limit be treated with the same probability?
A proper way to handle the model probability distribution near the boundary of a MG should be closely related to our understanding of the origin of MGs.
If a MG was formed in a loosely bound environment at birth without any noticeable over-density of their members against field stars, a uniform probability distribution model, i.e., assigning the same membership probability for all candidate members within the MG boundary, makes more sense.
However, if a MG was formed in a more gravitationally bound environment with a central concentration of members, a Gaussian probability distribution model may be more appropriate.
Between uniform or Gaussian distribution models in $XYZ$, to investigate which model is more appropriate to represent the actual distribution of MG members, one has to check the existence of an over-dense region in the spatial distribution of the models.
It is difficult to scrutinize any over-density in an already dispersed low-density group of stars.
Therefore, checking such an over-density in the back-traced positions in time is easier than a case of using the current positions.
However, the propagation of errors backward in time makes such analyses very difficult.
For given current knowledge and data of young MGs in the solar neighborhood, it is very difficult to distinguish between these two models.

Therefore, the virtue of using a uniform or a Gaussian $XYZ$ distribution model needs to be evaluated by the current distribution of members.
Fig.~\ref{fig4} shows the normalized number density of BPMG members (stars in the \listexc) in each $X$, $Y$, and $Z$.
For eliminating the binning effect in the investigation of the distributions, a kernel density estimation (KDE) was applied to these data.
It is well known that KDE underestimates the density near the boundary.
After correcting the boundary effect by truncating the kernel at the outermost boundaries (minimum and maximum values for each $X$, $Y$, and $Z$), $X$ and $Y$ values do not appear to be centrally concentrated.
However, $Z$ values appear to be more centrally concentrated.
It is interesting to note that BPMG members are concentrated in the $Z$-direction.
Because of the vertical structure of Milky Way with the scale height of about 300 pc, a certain degree of concentration of stars is expected but not at this level.
The peak of the $Z$-concentration is around $-$15 pc which may be the manifestation of the Sun being located 10$-$30 pc above the Galactic plane \citep{hum95, jos07}.

As shown in Fig.~\ref{fig4}, the difference in $XYZ$ distribution models is not that significant and the best model seems to be a combined one (uniform in $X$ \& $Y$ and Gaussian in $Z$).
If a future survey for BPMG members especially close to the known BPMG boundary is carried out, such a combined $XYZ$ distribution model is recommended.
In this paper, we select a version of the uniform distribution model in all three ($XYZ$) directions for simplicity.

We compared results from two different distribution models: Gaussian and uniform models (in Table~\ref{tab2}, \caseii\ and \caseiii, respectively).
Throughout the paper, the distribution function of BPMG members in $UVW$ is assumed to be a Gaussian similar to the treatment in BANYAN II.

\caseiiiex\ and \caseiiiin\ adopt members in the \listexcinc, to construct uniform $XYZ$ models for BPMG utilizing a minimum volume enclosing ellipsoid (MVEE) algorithm \citep{kum05}.
The model parameters for these two cases are presented in Table~\ref{tab3} (\unifexcl\ and \unifincl),  
 and the distribution of one model (\unifexcl) is shown in Fig.~\ref{fig5}.
The uniform membership probability functions fitted as MVEEs would be step functions; a constant within each ellipsoid, but zero outside of the ellipsoid.
To consider candidate members sitting barely outside of the distance limit of known members, we increase the lengths of principal axes of the MVEEs by 1.2 times.
Fig.~\ref{fig3} shows that probabilities from these two uniform models drop to zero above a certain distance, while those from Gaussian models never reach zero.

\begin{landscape}
\begin{table}
   \caption{Model parameters for BPMG. Parameters for other MGs (i.e., TWA, Tuc-Hor, AB Doradus, Columba, Carina, and Argus) are  taken from BANYAN II \citep{gag14}.}
   \label{tab3}
   \begin{threeparttable}
   \begin{tabular}{crrrrrrrrrrrrrrrrrr}
   \hline
   Name & $X$ & $Y$ & $Z$ & $\sigma_{X}$ & $\sigma_{Y}$ & $\sigma_{Z}$ & $\phi_{XYZ}$ & $\theta_{XYZ}$ & $\psi_{XYZ}$ & $U$ & $V$ & $W$ & $\sigma_{U}$ & $\sigma_{V}$ & $\sigma_{W}$ & $\phi_{UVW}$ & $\theta_{UVW}$ & $\psi_{UVW}$ \\ 
   & (pc) & (pc) & (pc) &  (pc) & (pc) & (pc) & ($^\circ$) & ($^\circ$) & ($^\circ$) & (\kms)  & (\kms) &  (\kms) &  (\kms) &  (\kms) &  (\kms) &  ($^\circ$) & ($^\circ$) & ($^\circ$)    \\
   \hline
   \multicolumn{19}{c}{Model parameters from BANYAN II} \\ \hline
   BPMG$_{\rm BANYANII}$ & 7.6 & -3.5 & -14.5 & 8.2 & 13.5 & 30.7 & -90.2 & 65.1 & -77.9& -11.0 & -15.6 & -9.2 & 1.4 & 1.7 & 2.5 & -113.0 & -70.3 & 76.6\\ \hline
\multicolumn{19}{c}{Model parameters (Section 2)} \\ \hline
\excl\tnote{a}& 9.4  & -5.6 & -13.5 & 7.8 & 13.5 & 29.1 & 88.7 & -62.1 & 78.6     & -10.7  & -16.0 & -9.3 & 1.4 & 1.5 & 2.4 & -105.9  & -51.2 & 84.5 \\
\incl\tnote{b}&  7.7 & -5.2 & -12.8 & 8.3 & 13.6 & 29.3 & 86.9 & -60.5 & 80.0       & -10.7  & -16.0 & -9.2 & 1.3 & 1.6 & 2.4 & -107.9 & -49.3 & 81.0 \\
\unifexcl\tnote{a}& 19.7 & -2.8 & -14.4 & 22.0 & 32.5 & 67.0 & -24.9 & 77.2 &-18.4 & \multicolumn{9}{c}{same to \excl} \\
\unifincl\tnote{b}& 13.1 & -3.9 & -10.0 & 29.1 & 33.2 & 68.5 & -93.6 & 62.1 & 86.8 & \multicolumn{9}{c}{same to \incl} \\ \hline
\multicolumn{19}{c}{Revised model parameters using the confirmed member list of BPMG in Table 5 (Section 3)} \\ \hline
BPMG$\rm_{revised}$& 13.4 & -3.4 & -18.1 & 19.1 & 32.0 & 71.2  & -71.6 & 74.8 & 111.2 & -10.4 & -15.9 & -9.1 & 1.2 & 1.4 & 2.2 & -93.1 & -45.5 & -81.5 \\
\hline
\end{tabular}
\begin{tablenotes}
   \item[a] The properties are obtained from the \listexc.
   \item[b] The properties are obtained from the \listinc.
   \item Notes. 
   \item 1. $X, Y, Z$ and $U, V, W$ are central positions of the ellipsoidal models.
  \item 2. $\sigma_X$, $\sigma_Y$, $\sigma_Z$, and $\sigma_U$, $\sigma_V$, $\sigma_W$ are the lengths of semi-major axes in the direction of the principal axes.  
  Values from uniform distribution models ($XYZ$ models for \unifexcl, \unifincl, and BPMG$\rm_{revised}$) are the lengths of semi-principal axes of the minimum volume enclosing ellipsoid (MVEE), while other values are 1-$\sigma$ lengths from the Gaussian models.
The $\sigma$ values from MVEE and Gaussian model are not comparable because all members should be enclosed within the $\sigma$ in MVEE, while a large portion of members ($\sim$80 per cent) are located outside of 1-$\sigma$ in the Gaussian model.
In the 3D Gaussian distribution, 20, 74, and 97 per cent of data are within 1, 2, and 3-$\sigma$ boundaries, respectively.
 \item 3. $\phi$, $\theta$, $\psi$ are Euler rotational angles of the ellipsoids in degrees.
 \end{tablenotes}
\end{threeparttable}
\end{table}

\begin{table}
   \caption{Group properties of the field star model in $UVW$ in this study.}
   \label{tab4}
   \centering
   \begin{threeparttable}
   \begin{tabular}{crrrrrrrrrr}
   \hline 
Name & $U$ & $V$ & $W$ & $\sigma_{U}$ & $\sigma_{V}$ & $\sigma_{W}$ & $\phi$ & $\theta$ & $\psi$ & Weight\tnote{a} \\
& (\kms)  & (\kms) &  (\kms) &  (\kms) &  (\kms) &  (\kms) &  ($^\circ$) & ($^\circ$) & ($^\circ$) \\
\hline
 FLD1 & -17.1 & -18.0 & -6.3 & 6.8 & 8.1 & 14.2 & 48.5 & -86.2 & -56.2 & 0.29\\
FLD2 & -32.0 & -16.2 & -8.1 & 15.8 & 16.5 & 19.4 & -54.7 & -80.7 & 112.6 & 0.26\\ 
FLD3 & 2.4    & -0.8   & -6.7 & 8.8   & 9.8   & 13.9  & 88.8 & 1.2 & -97.4 & 0.24\\
FLD4 & 22.1 & -16.8 & -9.5 & 16.1 & 16.8 & 21.3  & 110.9 & -67.8 & 104.1 &0.21\\
\hline 
\end{tabular}
\begin{tablenotes}
   \item[a] The relative number density.
\end{tablenotes}
\end{threeparttable}
\end{table}
\end{landscape}

\clearpage

\subsubsection{Distribution models of field stars}
\begin{figure*}
\includegraphics[width=0.99\columnwidth]{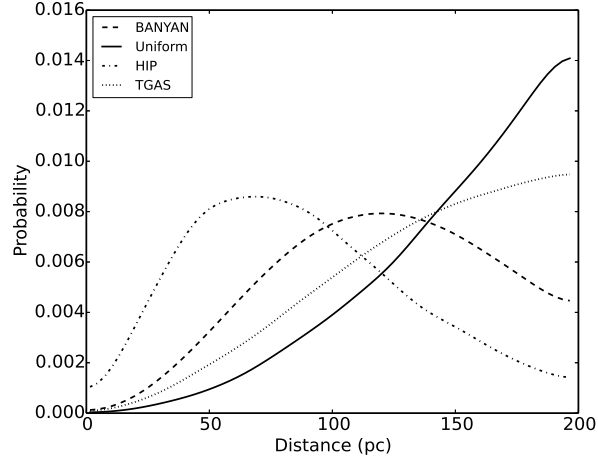}
\caption{Prior PDFs of distance for field star models used in BANYAN II (dashed line) and in this study (solid line).
BANYAN II includes two field star models (young and old), but their distance PDFs are similar and we present the old one.
Data from $Hipparcos$ (dot-dashed line) and TGAS (dotted line) catalogues are presented for comparisons.
Probabilities are normalized to $Hipparcos$ at 30 pc that would complete down to early M type stars.
We note that $Hipparcos$ data especially suffers from the limited survey depth, causing an apparent peak around $\sim$60 pc.}
\label{fig7}
\end{figure*}

Similar to the MG properties, field star properties can be acquired by finding the best-fit distribution model to the actual distribution of field stars in $XYZ$ and $UVW$.
In $XYZ$, we assumed that stars are uniformly distributed although this uniform spherical distribution may not be perfect in the Z direction at large distance because the scale height of the Galactic disk is about 300 pc.
This uniform field star model in $XYZ$ explains the actual distribution of nearby stars better than the case of BANYAN II.
Utilizing the Besan\c{c}on galaxy model \citep{rob03, rob12}, \citet{gag14} created a field star model by fitting the $XYZ$ distribution of all field stars inside of 200 pc with a single Gaussian model for the young ($<$1 Gyr) and old ($\ge$1 Gyr) field stars.
The field star model used in BANYAN II predicts a high concentration of field stars at $\sim$120 pc, 
while the largest stellar catalogue with measured parallactic distances at present ($Tycho-Gaia$ Astrometric Solution (TGAS); Michalik, Lindegren \& Hobbs  2015) shows no such concentration (Fig.~\ref{fig7}), which supports the uniform field star distribution in $XYZ$.

\begin{figure*}
\includegraphics[width=0.9\linewidth]{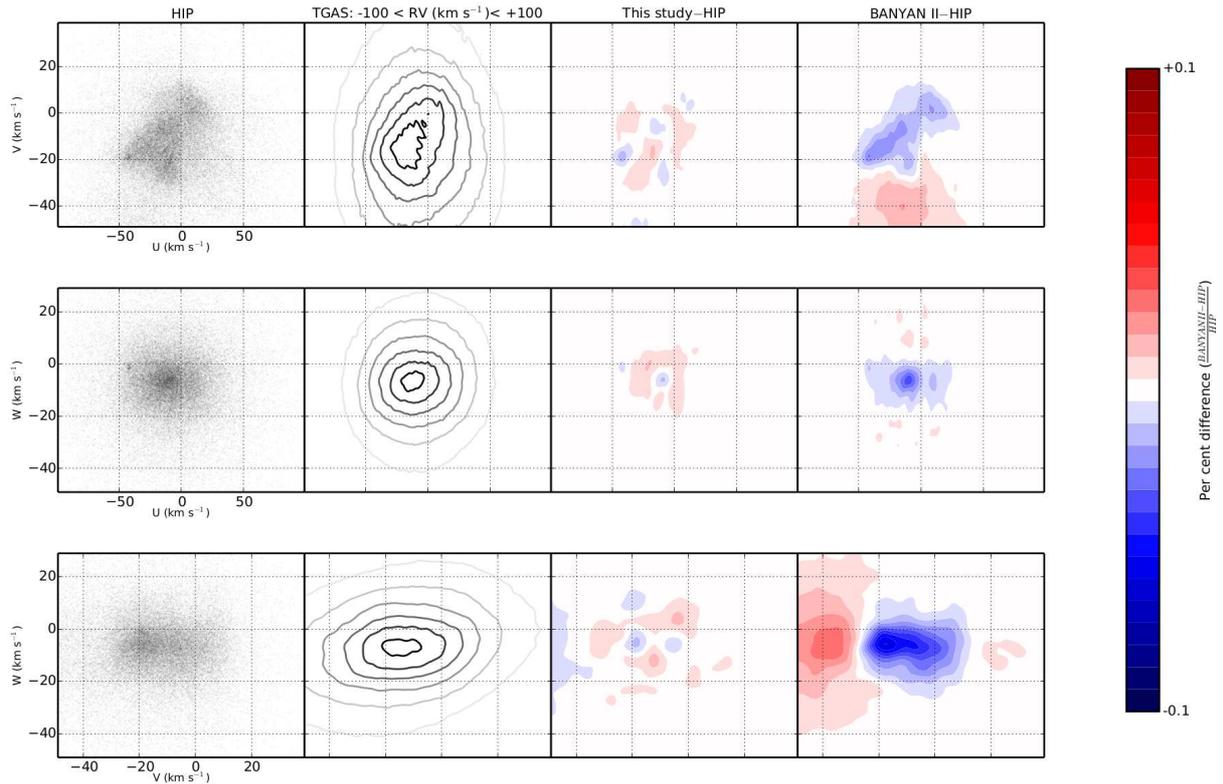}
\caption{Stellar distribution in $UVW$. First column represents the distribution of $Hipparcos$ stars with measured RV ($\sim$50,000 stars, \citet{and12}; references  therein).
Second column shows density maps of TGAS stars of possible $UVW$ values over the RV range of $-$100 to $+$100 \kms.
Residual contour plots at 3rd and 4th columns represent the differences between $Hipparcos$ stars and simulated stars generated by field star model from this study (3rd column), and those from BANYAN II (4th column).
Colors correspond to the colorbar, which scales from the minimum to the maximum values of the differences, that appear in the BANYAN II$-$HIP map on the $VW$ plane.}
\label{fig8}
\end{figure*}

\begin{figure*}
  \includegraphics[width=\textwidth]{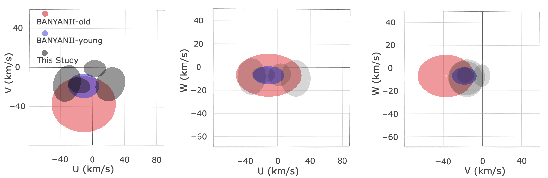}
  \caption{Field star distribution models from this study (grey) and those from BANYAN II (red and blue represent old and young field stars, respectively.).
  Plotted are 1-$\sigma$ ellipses.}
 \label{fig9}
\end{figure*}

The field star distribution in $UVW$ was examined using $Hipparcos$ and TGAS catalogues.
To examine the distribution of stars in $UVW$, one has to know six parameters: RA, Dec., distance, proper motions in RA and Dec., and RV.
Aproximately 50,000 stars in the $Hipparcos$ catalogue have all six parameters, and they clearly show 3-4 subgroups in $UVW$ (Fig.~\ref{fig8}).
By applying a wavelet analysis technique on the 2D $UVW$ plots, Skuljan, Hearnshaw \& Cottrell (1999) and Chereul, Cr\`{e}z\`{e} \& Bienaym\`{e} (1997) identified $\sim$4 kinematic clusters of nearby field stars in the $Hipparcos$ catalogue.
Because the $Hipparcos$ catalogue is limited to stars mostly within $<$100 pc from the Sun, one has to use a larger kinematic catalogue to check if the apparent 4 kinematic clusters of nearby stars exist beyond 100 pc.
The TGAS catalogue provides 5 parameters for $\sim$2 million stars, and the only missing parameter to calculate $UVW$ is RV.
Since almost all nearby stars do not travel faster than $\sim$100 \kms\ with respect to the Sun, we can assume that most TGAS stars have RVs in the range of $-$100 and $+$100 \kms.
Calculating possible $UVW$s over the range of these RVs, we can examine the kinematic clustering of field stars using a much larger sample of stars ($\sim$2 million) than the $Hipparcos$ data ($\sim$50,000).
Even though the stellar distribution becomes diluted because of the $\pm$100 \kms\ RV range instead of a single value of RV for a star, the kinematic clustering of TGAS stars looks similar to that of $Hipparcos$ stars (first and second columns in Fig.~\ref{fig8}).

Adopting these two possible improvements described above, we created a new field star model.
The new field star model in $UVW$ is obtained from the best-fit Gaussian ellipsoidal models of $Hipparcos$ stars (Table~\ref{tab4} and 2D projections in Fig.~\ref{fig9}), and the model in $XYZ$ is a uniform distribution within a sphere of 200 pc in radius.
As can be seen in Fig.~\ref{fig7}, the distance distribution of the field star model in BANYAN II peaks at $\sim$120 pc, while the PDF of the new field star model expects more stars at larger distance.
A field star distribution model is involved in the overall normalisation of the calculation, therefore, this uniform field star distribution model would increase the MG membership probability for nearby MG members;
however, it would decrease the membership probability of more distant MG candidate members.
Fig.~\ref{fig8} (3rd and 4th columns) compares the field star $UVW$ model used in BANYAN II to the new model, showing that the latter fits the $Hipparcos$ more closely.

\section{Results}
Using the updated models mentioned in the previous section, we now calculate membership probabilities of all BPMG candidate members available from SIMBAD \citep{zuc01a, son03, zuc04, moo06, tor06, tor08,  lep09, tei09, sch10, sch12a, sch12b, shk12, mal13, moo13, mal14, bes15, gag15a, gag15b}.
The result shows a sensitive dependence on models.
Based on the calculated membership probabilities, we updated the list of BPMG members adding 12 new bona fide members.
With this updated membership list, we revised the BPMG model as described below.

\subsection{Effects of improved models}
\begin{figure*}
      \includegraphics[width=0.9\linewidth]{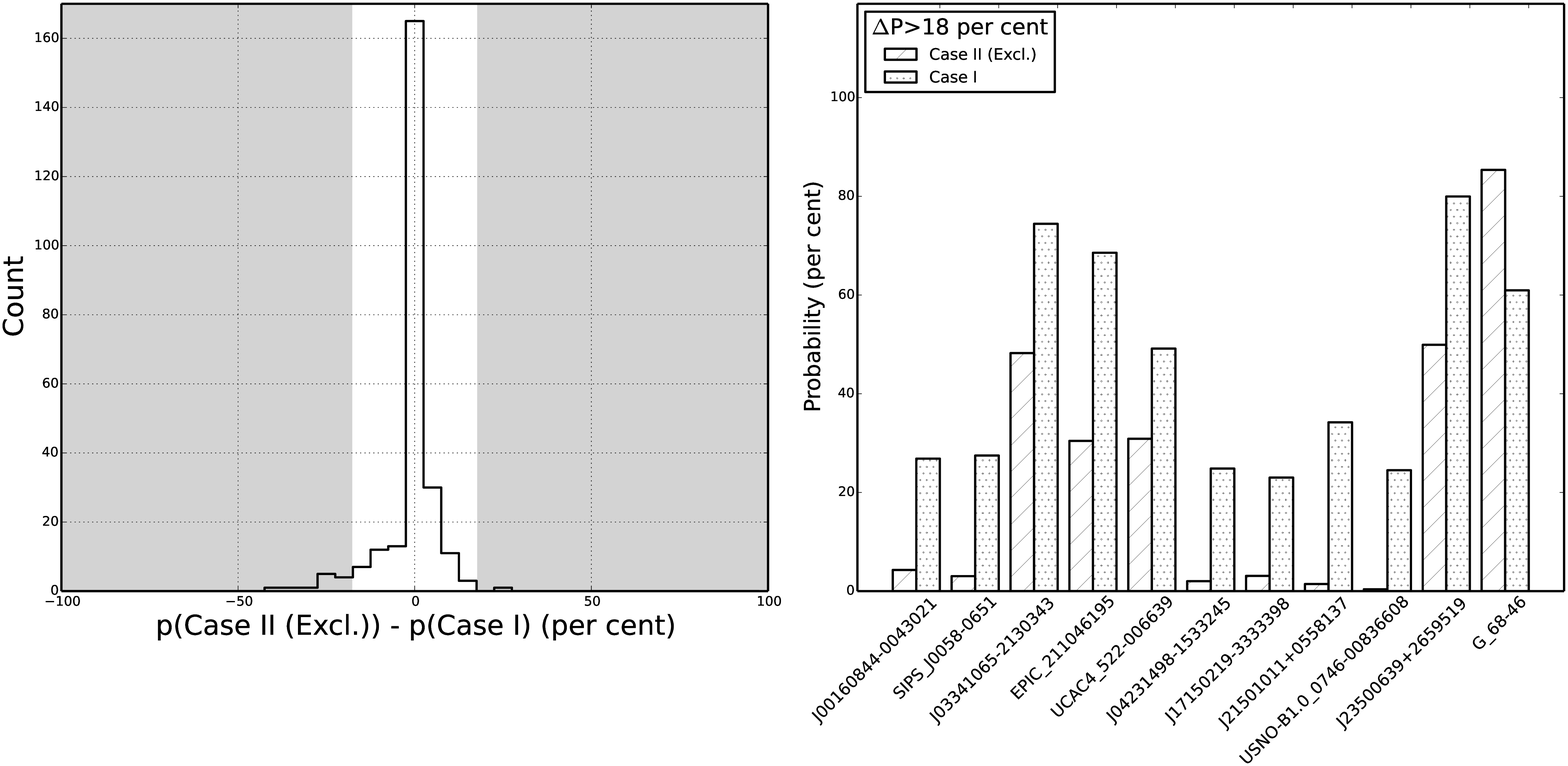}
   \caption{Effects of the BPMG member selection (BANYAN II (\casei) versus \listexc\ (\caseii)). 
Left panel shows a histogram of the membership probability differences between these two cases.
Stars showing large differences in the membership probabilities ($\Delta p >$ 18 per cent, grey area in the left panel) are presented in the right panel.
  Test stars are from the \listthree.}
    \label{fig10}
\end{figure*}

\begin{figure*}
      \includegraphics[width=0.9\linewidth]{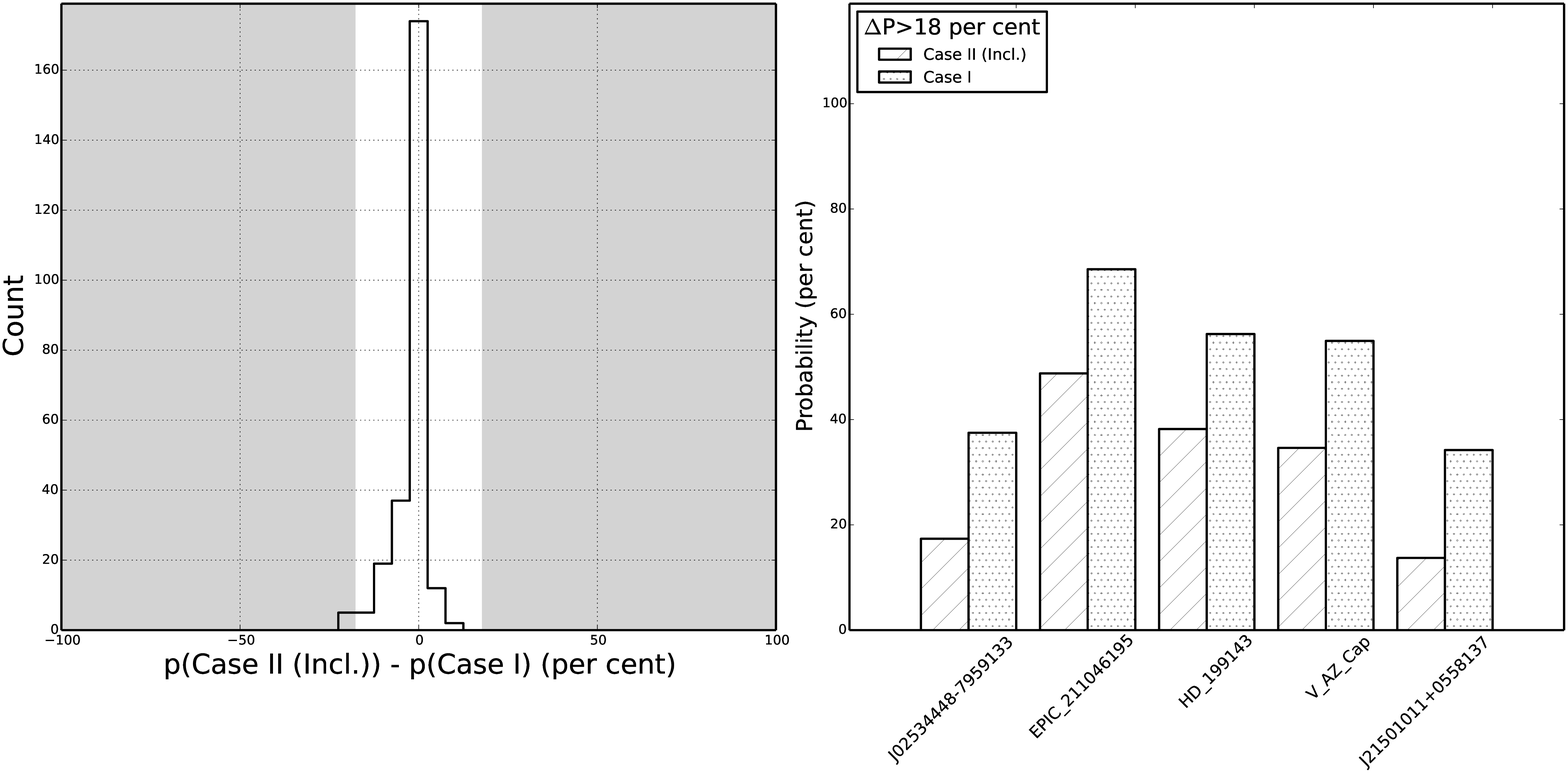}
   \caption{Effects of the BPMG member selection (BANYAN II (\casei) versus \listinc\ (\caseii)). 
   Left panel shows a histogram of the membership probability differences between these two cases.
Stars showing large differences in the membership probabilities ($\Delta p >$ 18 per cent, grey area in the left panel) are presented in the right panel.
 Test stars are from the \listthree.}
   \label{fig11}
\end{figure*}

\begin{figure*}
      \includegraphics[width=0.9\linewidth]{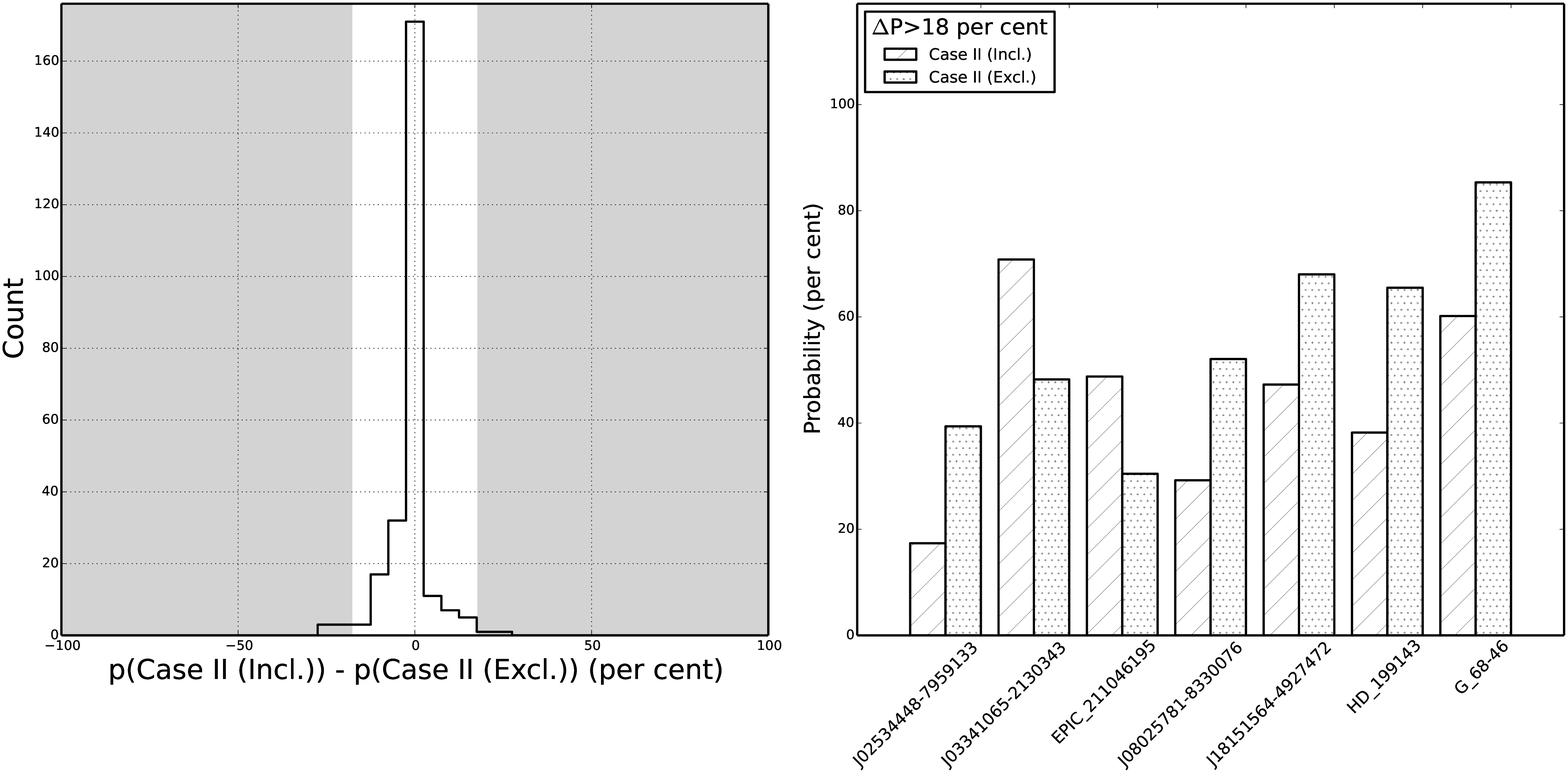}
   \caption{Effects of the BPMG member selection (\listinc\ versus \listinc).
  Left panel shows a histogram of the membership probability differences between these two cases.
Stars showing large differences in the membership probabilities ($\Delta p >$ 18 per cent, grey area in the left panel) are presented in the right panel.
 Test stars are from the \listthree.}
   \label{fig12}
\end{figure*}

We examine the effects of membership by comparing cases that adopt the \listexcinc\ (Table~\ref{tab2}). 
Since BANYAN II used a Gaussian $XYZ$ distribution (\casei), we compare \casei, \caseiiex\ and \caseiiin, using a Gaussian distribution for members in $XYZ$.
Since the BPMG model based on \listinc--(\caseiiin)--allows more marginal members to start with, this flexibility would increase the membership probability of a marginal member, while the model from \listexc--(\caseiiex)--would decrease the probability.
Figs.~\ref{fig10}$-$~\ref{fig12} compare membership probabilities from \casei\ and those from \caseii.
Overall, the effects of the member list appear to be small (less than 5 per cent of  test stars having a probability difference of greater than 20 per cent), and it is likely due to the similarities of the model extents (see $\sigma$ values for \excl, \incl, BPMG$\rm_{BANYANII}$ in Table~\ref{tab3}).
However, for a few stars, the effect of the member list is significant causing the membership probability changes up to $\sim$40 per cent.

\begin{figure*}
     \includegraphics[width=0.9\linewidth]{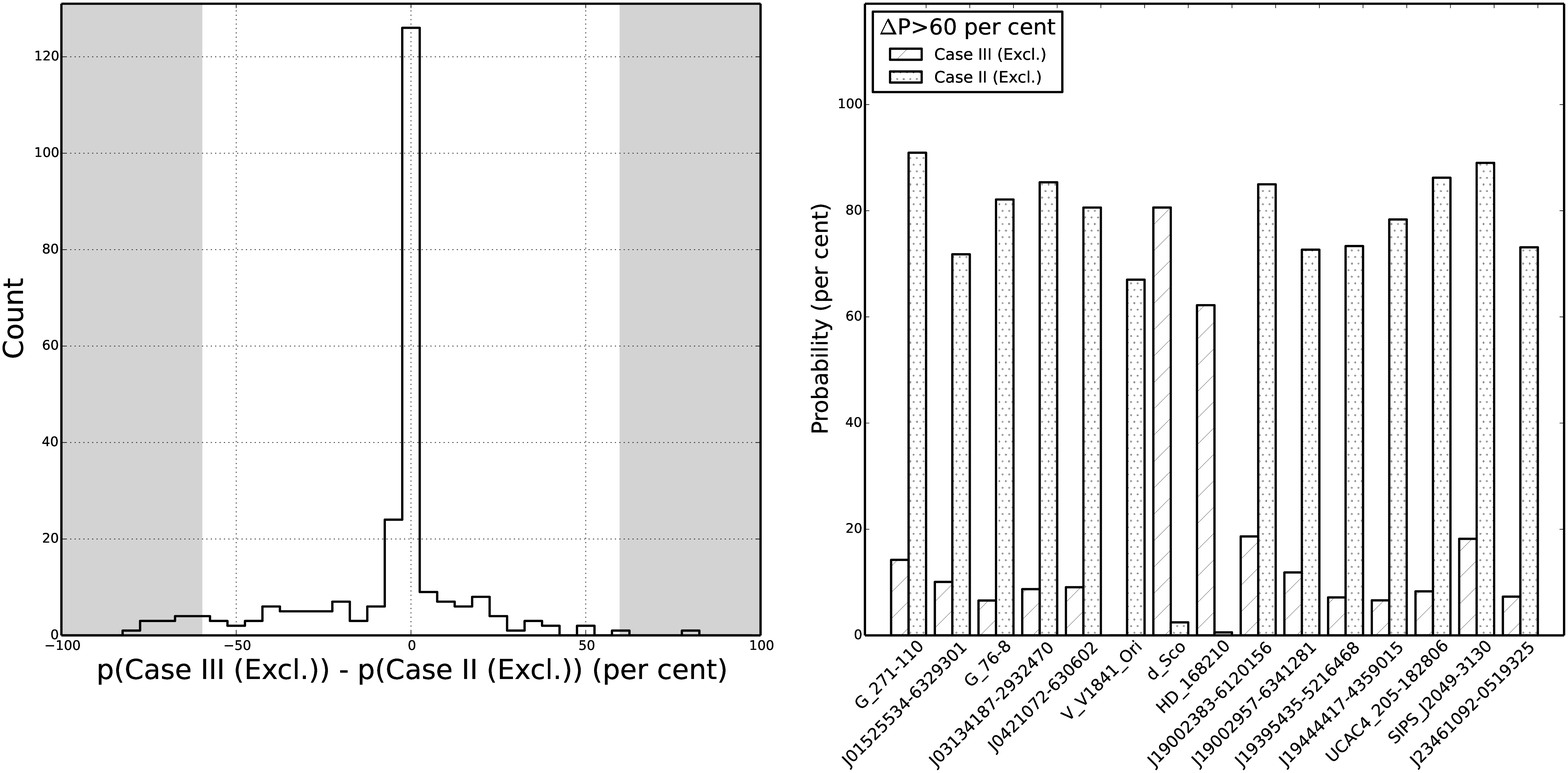}
   \caption{Effects of the distribution model of BPMG members (a Gaussian (\caseiiex) versus a uniform (\caseiiiex) distribution in $XYZ$). 
   Left panel shows a histogram of the membership probability differences between these two cases.
Stars showing large differences in the membership probabilities ($\Delta p >$ 60 per cent, grey area in the left panel) are presented in the right panel.
 Test stars are from the \listthree.}
   \label{fig13}
\end{figure*}

\begin{figure*}
     \includegraphics[width=0.9\linewidth]{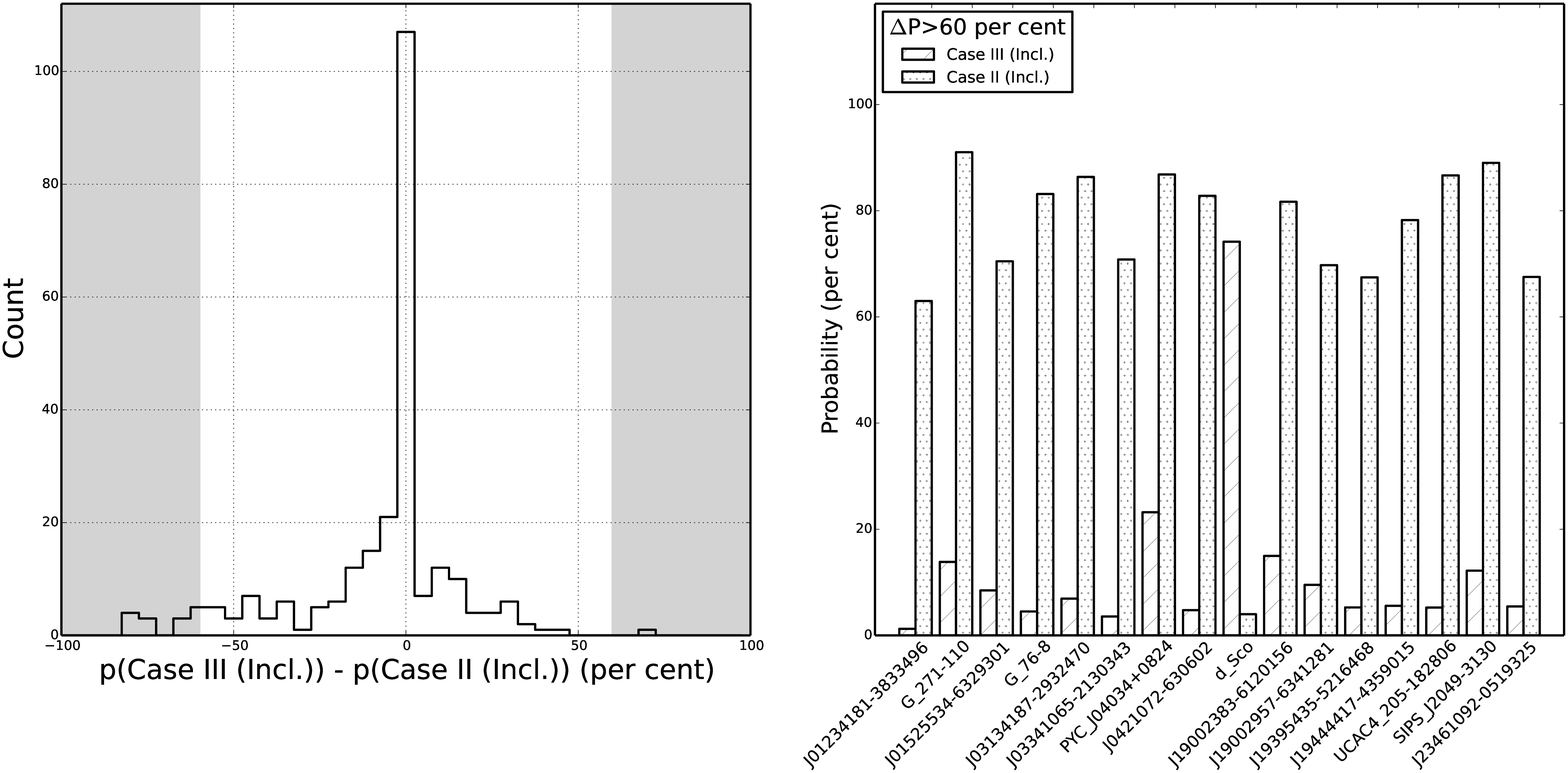}
   \caption{Effects of the distribution model of BPMG members (a Gaussian (\caseiiin) versus uniform (\caseiiiin) distribution in $XYZ$). 
  Left panel shows a histogram of the membership probability differences between these two cases.
Stars showing large differences in the membership probabilities ($\Delta p >$ 60 per cent, grey area in the left panel) are presented in the right panel.
Test stars are from the \listthree.}
   \label{fig14}
\end{figure*}

Figs.~\ref{fig13} and ~\ref{fig14} compare the membership probabilities from cases assuming Gaussian distribution (\caseii) or uniform distribution (\caseiii) of members in $XYZ$.
As expected, the Gaussian distribution models decrease membership probabilities of candidate members located near the boundary.
The membership probabilities of HD 168210 and $\delta$ Sco were increased by $\sim$60 per cent under the uniform spatial distribution model.
These two stars are located far from the centre of the BPMG model ($\sim$70 and $\sim$40 pc, respectively).
Stars relatively close to the centre, such as G76-8 ($\sim$20 pc) on the other hand, have larger membership probabilities under the Gaussian distribution model ($>$80 per cent in \caseii\ versus $<$10 per cent in \caseiii).
This difference can be important in investigating the membership status of candidate members around or beyond the assumed initial MG boundary.

\begin{figure*}
      \includegraphics[width=0.9\linewidth]{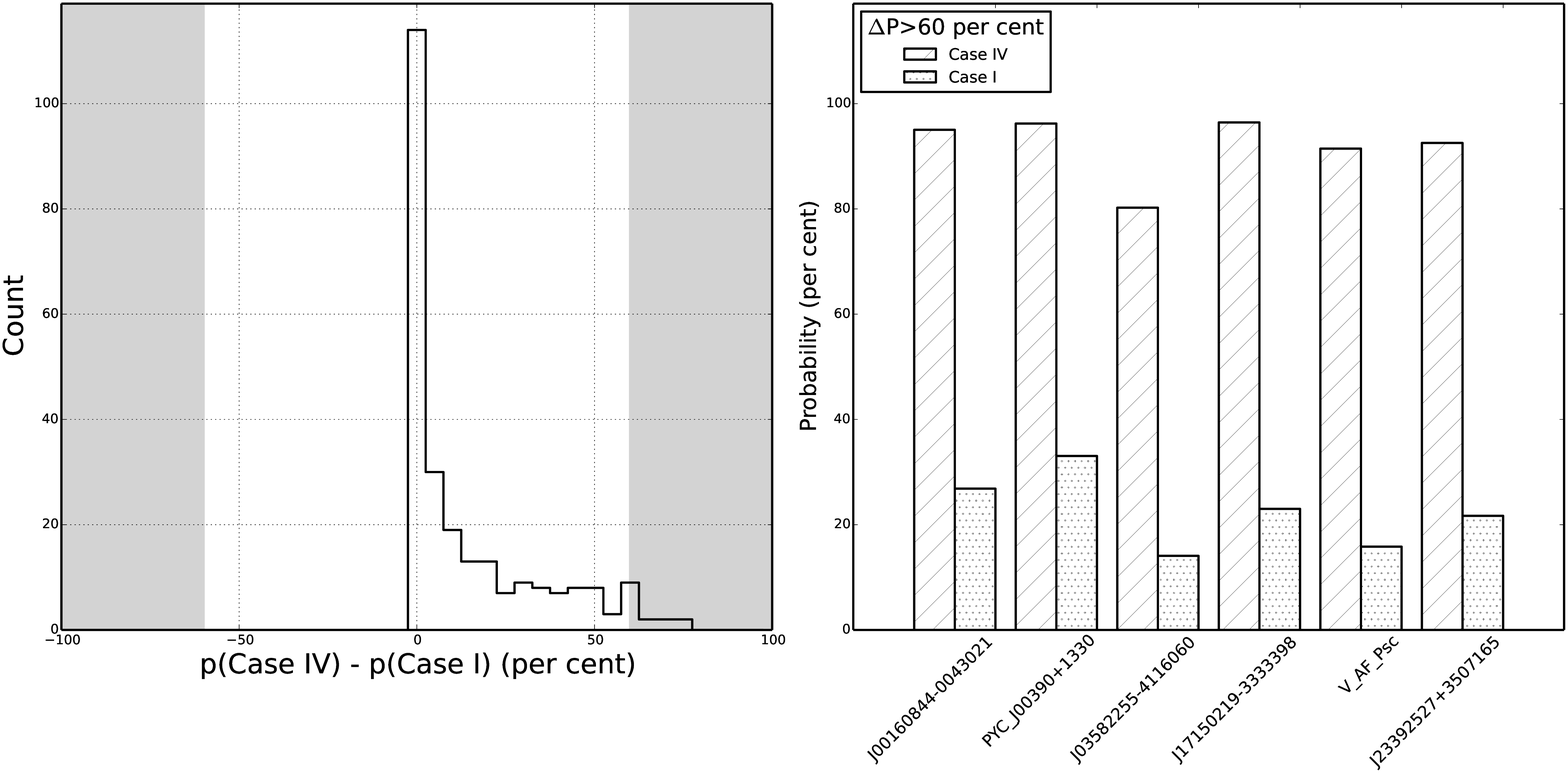}
  \caption{Effects of the field star model (BANYAN II (\casei) versus a new field star model  (\caseiv)).
  Left panel shows a histogram of the membership probability differences between these two cases.
Stars showing large differences in the membership probabilities ($\Delta p >$ 60 per cent, grey area in the left panel) are presented in the right panel.
 Test stars are from the the \listthree.}
   \label{fig15}
\end{figure*}

\begin{figure*}
   \includegraphics[width=0.9\linewidth]{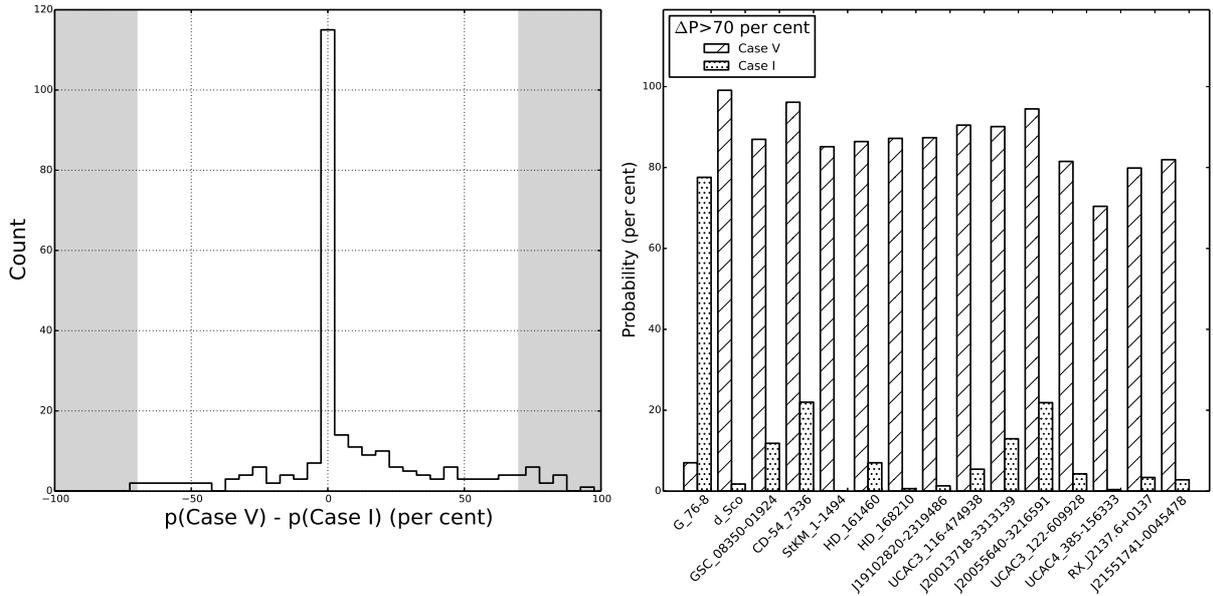}
\caption{A comparison of membership probabilities from \casev\ (using all updated models simultaneously) and those from \casei\ (BANYAN II).
   Left panel shows a histogram of the membership probability differences between these two cases.
Stars showing large differences in the membership probabilities ($\Delta p >$ 70 per cent, grey area in the left panel) are presented in the right panel.
 Test stars are from the \listthree.}
\label{fig16}
\end{figure*}

\begin{figure}
  \includegraphics[width=0.9\columnwidth]{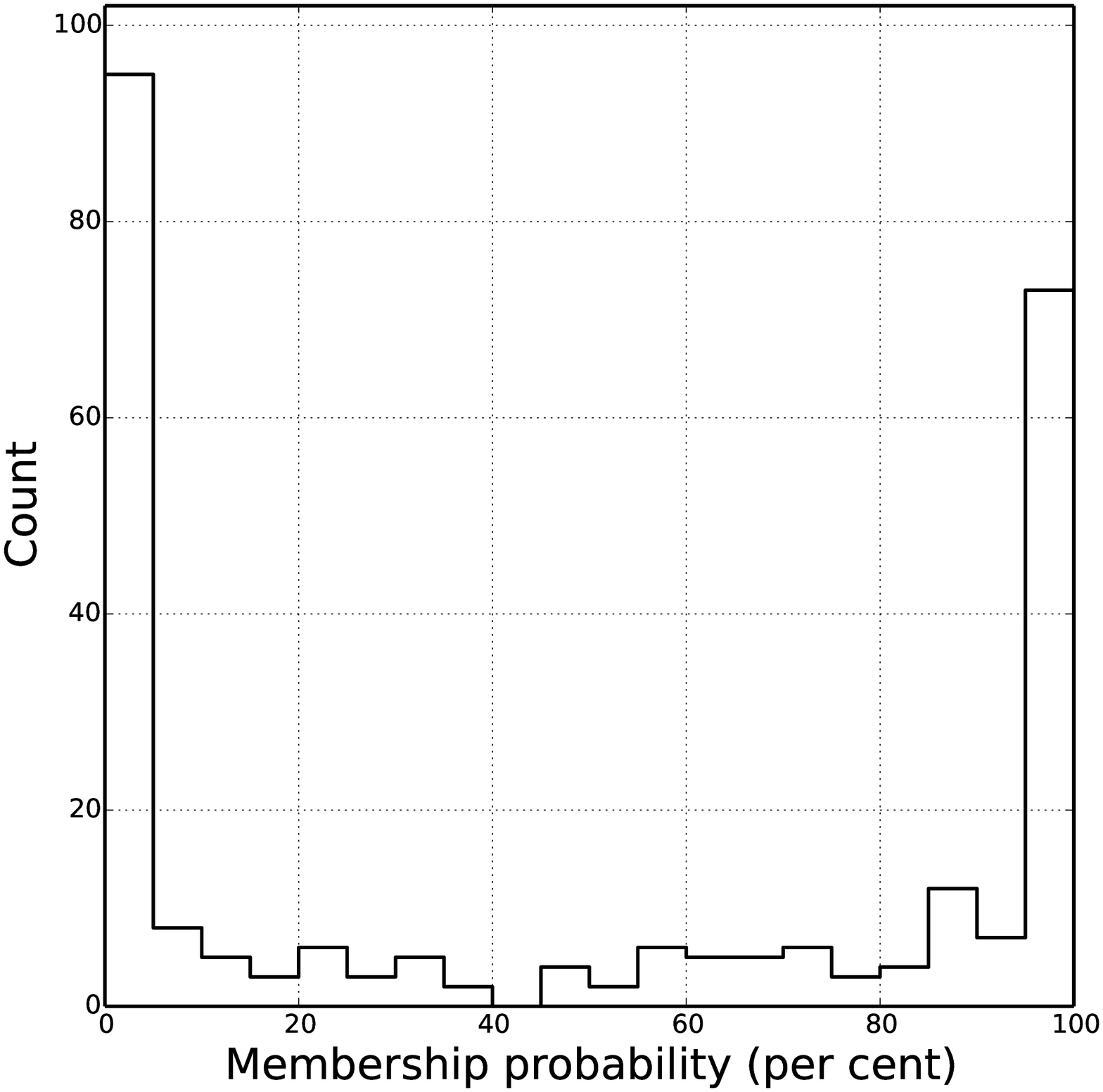}   
\caption{Membership probabilities for stars from the \listthree\ using all updated models (\casev).}
\label{fig17}
\end{figure}

Because there are many more old field stars than MG members in a given range of $XYZUVW$, a small change in the $XYZUVW$ distribution model of field stars can significantly affect membership probabilities of candidate MG members.
Fig.~\ref{fig15} shows the effect of the field star model by comparing membership probabilities from \caseiv\ and those from \casei.
Membership probabilities tend to increase under the new model of field star distribution (\caseiv), by up to 80 per cent.
\caseiv\ assumes a uniform field star distribution in $XYZ$, expecting a smaller field star number density, within $\sim$140 pc, compared to BANYAN II (Fig.~\ref{fig7}).
Since almost all known BPMG candidate members are located within $\sim$100 pc, membership probabilities from \caseiv\ generally increase compared to those from \casei.

We have shown that each case listed in Table~\ref{tab2} significantly affects the membership probability.
Using all of these updated models simultaneously (\casev), we calculated membership probabilities of stars in the \listthree.
These membership probabilities are compared to those from \casei\ (Fig.~\ref{fig16}),  showing a significant difference in membership probability.
About 40 stars show changed membership probabilities larger than $\sim$50 per cent compared to values from \casei.
The majority of stars in \listthree\ ($\sim$60 per cent) have membership probabilities of less than 50 per cent, implying the high contamination rate of false members in the \listthree\  (Fig.~\ref{fig17}). 

\clearpage

\subsection{Membership assessment and a revised BPMG model based on the improvement}

We can reconstruct a new list of bona fide members of BPMG based on our updated scheme of using several updated models simultaneously (the updated field star model, uniform $XYZ$ distribution of BPMG members, and either the {\it exclusive} or {\it inclusive} list of initial adopted members).
When this new scheme was applied to 275 candidate members from the \listthree, only about 40 per cent stars are believed to be kinematically associated with BPMG (p $>$50 per cent).
A more stringent selection of candidate members (p$>$80 per cent) indicates only one third of the suggested members can be retained.
Moreover, a kinematic similarity is not a sufficient condition for being a true member because of the large contamination of field stars with similar kinematics.
To be deemed as a bona fide member, any survived kinematic candidates should also show a clear signs of youth ($\lesssim$25 Myr).
In this section, we discuss details on how we reject or include a particular star in a list of updated bona fide members.

\subsubsection{Method of rejection}
Stars with small kinematic membership probability ($<$80 per cent) or lacking clear signs of youth (age $\lesssim$25 Myr) were rejected.
We present five cases of rejected stars from the \listone.

\paragraph{HIP 11360}
HIP 11360 was initially suggested as a BPMG member by \citet{moo06}, while \citet{mal13} suggested it as a Columba member.
Although the star shows strong Li absorption, it should not be considered as a BPMG member due to the low membership probability (0 per cent) in agreement with \citet{mal13}.
Instead, we suggest that HIP 11360 is a Tuc-Hor member because of the large membership probability (p(Tuc-Hor)$\sim$100 per cent).

\paragraph{HIP 50156}
HIP 50156 was initially proposed as a member of BPMG by \citet{sch12a}, while \citet{mal13} suggested that it is likely to be a member of Columba.
In spite of showing unambiguous youth ($\lesssim$25 Myr; based on X-ray luminosity, photometric magnitude, and NUV-excess), this star should not be considered as a BPMG member due to the low membership probability (0 per cent).

\paragraph{2M J06085283-2753583}
2M 06085283-2753583 was initially suggested as a BPMG member by Rice, Faherty \& Cruz (2010). 
 However, this star has a low kinematic membership probability (1 per cent).
In addition, unambiguous young age cannot be demonstrated based on a CMD position (e.g., $V-K$ versus $M_K$) because of the large model uncertainty of PMS evolution models at this young age and low mass range.
Furthermore, an empirical comparison against other known young stars is implausible because of lacking comparison stars with demonstrated youth.

\paragraph{$\eta$Tel A\&B}
$\eta$ Tel A\&B were originally proposed as members of Tuc-Hor by \citet{zuc00}, but their memberships were later revised to BPMG by \citet{zuc01a}.
$\eta$ Tel A shows unambiguous youth on the CMD, but it should not be considered a BPMG member due to the low kinematic membership probability (4 per cent).
Its companion, $\eta$ Tel B, has a large membership probability ($\sim$90 per cent), which is obtained by marginalising over RV.
In order for $\eta$ Tel B to be a BPMG member, its predicted RV must be $\sim$1 \kms, which is significantly different from the consistently measured value of 12$-$14 \kms\ for $\eta$ Tel A \citep{cam28, wil53, eva64, eva67, gon06}.
Thus, these stars should not be considered as BPMG members.

\subsubsection{Method of inclusion}
In order to be considered as bona fide members, stars should show large kinematic membership probability ($>$80 per cent) and clear signs of age younger than or similar to 25 Myr.
39 stars in the \listone\ are confirmed as bona fide BPMG members.
They are listed in the first part of Table~\ref{tab5}.
Among these stars, HIP 86598 and HIP 89829 are located far from Earth compared to other members ($\sim$70$-$80 pc; other members are located within 50 pc).
HIP 86598 was suggested as a Upper Scorpius member by Song, Zuckerman \& Bessell (2012), while \citet{mal13} suggested this star as a BPMG member.
The other star, HIP 89829, was identified as a BPMG member by \citet{tor08} and \citet{mal13}.
Both HIP 86598 and HIP 89829 can be BPMG members because their Z position (about $-$10 pc) is in the range for BPMG members ($-$40 to $+$10 pc) rather than for stars of Upper Scorpius ($+$20 to $+$100 pc).
Their positions in $XYZ$ and/or $UVW$ appear to be close to the edge of the BPMG model ($X, Y, U$, and $V$ for HIP 86598, $X$ and $Z$ for HIP 89829).
Therefore, these stars can be treated marginal BPMG members.

In addition to these confirmed members in the \listone, we added 12 new members from the \listthree\ spanning spectral types K1 to M4.5.
These members have  kinematic membership probabilities larger than 95 per cent and clear signs of youth.
They are shown in Table~\ref{tab5} designated as newly confirmed members.
These 12 stars were suggested as BPMG candidate members in previous researches \citep{tor06, lep09, mes10, kis11, mal13, mal14, rie14}.
2MASS J21212873-6655063 was initially suggested as a member of Tuc-Hor by Zuckerman, Song \& Webb (2001b);
 however, subsequent works \citep{mal13, mal14} and the kinematic membership probability (100 per cent) based on the updated models in this study support that this star should be a bona fide member of BPMG.
In addition, CD-54 7336 and CD-31 16041 were initially proposed as members of Upper Scorpius by \citet{son12} based on the direction in the sky and large assumed distances.
Now, we also conclude that these two stars are members of BPMG instead of Upper Scorpius based on their updated membership probabilities (97 to 100 per cent), Z positions ($-$12 to $-$13 pc), and trigonometric distances ($\sim$50 to 70 pc).

\subsubsection{Ambiguous cases}
Among well-known members of BPMG (the \listone), 6 stars (HIP 10679, HIP 10680, HIP 21547, HIP 88726A and B, HIP 92024) are ambiguous in their membership determination because they show moderate signs of youth ($\lesssim$100 Myr but not $\lesssim$25 Myr).
We retained them in our bona fide member list for the lacking clear evidence of their non-memberships.

\subsubsection{Suggestion of probable members}
Finally, there are 17 highly probable BPMG candidate members.
We separately listed these stars in 3 groups in Table 5 according to age and data constraints.
Group 1 consists of stars showing unambiguous youth with large membership probabilities ($>$85 per cent), but lacking partial kinematic parameters such as distance or RV.
The 4 stars in this group were suggested as candidate members of BPMG in several studies \citep{zuc04, tor06, mes10, mal13, mal14}.
HD 161460 was suggested as a member of Upper Scorpius by \citet{son12}.
Group 2 contains stars having large membership probabilities ($>$80 per cent), but showing only moderate signs of youth and missing distance or RV.
Stars in this group were suggested as members of BPMG in multiple studies \citep{lep09, sch10, kis11, mal13, mal14, gag15a, gag15b}.
However, PYC J00390+1330 and UPM J1354-7121 were also suggested as AB Dorarus members by \citet{sch12a} and \citet{mal13}, respectively.

A single star in group 3, TYC 6872-1011-1, has the full 6 kinematic parameters showing unambiguous youth, but its kinematic probability is slightly low ($\sim$60 per cent).
This star was also suggested to be a BPMG member in several studies \citep{tor06, mes10, mal13, mal14}.

\subsubsection{Summary}
Among all 275 BPMG candidate members in the \listthree, 57 stars can be confirmed as bona fide BPMG members.
39 stars are from the \listone, and 12 stars are newly confirmed.
We additionally include traditional 6 BPMG members showing ambiguity in youth.
These stars should be removed in the future if they show clear evidence of non-memberships.
Five stars from the \listone\ are rejected, mainly due to updated low kinematic membership probabilities. 
We note that some of the false members were used in several age-related studies (e.g., absolute isochronal age scale; Bell, Mamajek \& Naylor (2015), lithium depletion boundary age; \citet{bin16}), which could have biased the results.

The list of updated bona fide members was utilized to revise a BPMG model (BPMG$\rm_{revised}$ in Table~\ref{tab3}), which, in turn, can be used in future searches for members based on new data from the $Gaia$ mission.

\begin{table*} 
\caption{The updated list of BPMG members.}
\label{tab5}
  \centering 
  \tiny 
  \setlength\tabcolsep{2pt}
 \begin{threeparttable} 
  \begin{tabular}{cccccccccccccccccccc} 
  \hline 
Name & SpT. & R.A. & Dec. & $\mu\rm_{\alpha}$ & $\mu\rm_{\delta}$ & $\mu$ & $\pi$ & $\pi$ & RV & RV & B-V & V-K & K & NUV & \lxlbol\ & Li\tnote{a} & Li & p(BPMG)\tnote{b} \\ 
   & & (hh:mm:ss) & (dd:mm:ss) & (mas yr$^{-1}$) & (mas yr$^{-1}$) &  Ref. & (mas) & Ref. & (\kms) & Ref. & (mag) & (mag) & (mag) & (mag) & & (m\AA) & Ref. & (\%) \\ \hline 
 \multicolumn{20}{c}{Confirmed Members from a Previously Known BPMG Member List (the \listone)} \\ \hline
                              HIP 560 &             F2IV &     00:06:50 &    -23:06:27 &    97.1      $\pm$     0.03 &   -47.3   $\pm$     0.02 & 5 &    25.2  $\pm$     0.4  & 5 &      6.5 $\pm$      3.5 & 7 &   0.38 &   0.93 &   5.24 &    $-$  &  -5.34 &  87 &  25  &   100 \\ 
             2MASS J01112542+1526214 &  M5 &     01:11:25 &    +15:26:21 &    192.0 $\pm$      8.0 &   -130.0 $\pm$      8.0 & 30 &    45.8  $\pm$     1.8 & 17 &      4.0 $\pm$      0.1 & 17 &   1.76 &   6.22 &   8.21 &  19.88 &  -3.00 & $-$ & $-$ &100 \\ 
              2MASS J01351393-0712517 &M4 &     01:35:13 &    -07:12:51 &     93.0 $\pm$      1.7 &    -48.0  $\pm$      2.2 & 30 &     25.9                           & 20 &      6.3 $\pm$      0.5 & 12 &   1.50 &   5.35 &   8.08 &  19.05 &  -2.56 & $-$ & $-$  &  99\\ 
                           HIP 10679 &              G2V &     02:17:24 &    +28:44:31 &    85.1 $\pm$     0.4  &   -70.8   $\pm$     0.3   & 5 &    25.5  $\pm$     0.2 & 5 &      5.7 $\pm$      0.3 & 33 &   0.62 &   1.50 &   6.26 &  12.94 &  -3.91 & 163 & 13,14,4,25,22 & 100\\ 
                          HIP 10680 &              F5V &     02:17:25 &    +28:44:43 &    87.1 $\pm$     0.2   &   -74.1  $\pm$     0.2  & 5 &     25.2 $\pm$     0.2 & 5 &      5.4 $\pm$     0.5 & 37 &   0.51 &   1.24 &   5.79 &  12.63 &  -4.18 & 132 & 13,14,4,25,22 &100 \\ 
                          HIP 11152 &             M3V &     02:23:26 &    +22:44:06 &    98.5  $\pm$     0.2   &   -112.5 $\pm$     0.1  & 5 &    36.9  $\pm$     0.3  & 5 &     10.4 $\pm$      2.0 & 18 &   1.56 &   3.93 &   7.35 &  17.77 &  -3.16 &   0 & 32 & 100\\ 
                       HIP 11437 B &               M0 &     02:27:28 &    +30:58:41 &     81.5 $\pm$      4.6 &    -69.1 $\pm$      3.2 & 30 &    24.3 $\pm$     0.2  & 5 &      4.7 $\pm$      1.3 & 22 &   1.50 &   4.63 &   7.92 &  18.98 &  -2.37 & 115 & 14,4,25,22 &100 \\ 
                      HIP 11437 A &               K8 &     02:27:29 &    +30:58:25 &     79.5 $\pm$     0.2   &   -72.1  $\pm$     0.1  & 5 &    24.3   $\pm$     0.2  & 5 &     6.5 $\pm$     0.4 & 27 &   1.33 &   2.99 &   7.08 &  17.64 &  -3.10 & 227 & 14,4,25,22 & 100\\ 
                        HIP 12545 AB &             K6V &     02:41:25 &    +05:59:19 &    79.5 $\pm$     3.1 &   -53.9  $\pm$     1.7  &27 &    23.8  $\pm$      1.5 & 27&     10.0   & 25 &   1.09 &   3.15 &   7.07 &  17.47 &  -2.94 & 433 &28,14,4,25,22 & 97 \\ 
             2MASS J03350208+2342356 & M8.5 &  03:35:02   &   +23:42:36  &   54$\pm$10          &     -56 $\pm$ 10          & 20 &   23.1                        & 20 &     15.5 $\pm$1.7 & 20 &         $-$ & $-$ &        11.26 & 22.12& $-$ & 615   & 20 & 85 \\
                            HIP 21547 &              F0V &     04:37:36 &    -02:28:25 &    44.2  $\pm$     0.4  &   -64.4   $\pm$     0.3  & 27 &    34.0  $\pm$     0.3  & 27 &     12.6 $\pm$      0.3 & 27 &   0.28 &   0.67 &   4.54 &    $-$ &    $-$  &   0 & 25&    94 \\ 
           GJ 3305 AB  &  M0V &     04:37:37 &    -02:29:28 &     45.9 $\pm$      1.3 &    -63.6 $\pm$      1.2 & 30 &    34.0  $\pm$     0.3  & 27 &    21.7$\pm$  0.3 & 20&   1.45 &   4.18 &   6.41 &   $-$ &  -2.52 & 99 & 14,25 &100 \\ 
                          HIP 23200 &             M0V &     04:59:34 &    +01:47:00 &    39.3  $\pm$     0.2  &   -95.0  $\pm$      0.1 &5  &    40.7  $\pm$     0.3 &5 &    19.3 $\pm$     0.2 & 34&   1.38 &   3.84 &   6.26 &    $-$ &  -3.05 & 270 & 2,25& 100\\ 
                           HIP 23309 &                M0 &     05:00:47 &    -57:15:26 &     35.3 $\pm$      0.1 &    74.1   $\pm$      0.1 & 5 &    36.9  $\pm$     0.3 &5 &     19.4 $\pm$      0.3 & 25&   1.38 &   3.73 &   6.24 &  17.51 &  -3.33 & 325 & 28,14,4,25,31 &100\\ 
                      HIP 23418       &              M3V &     05:01:58 &    +09:58:60 &    12.1  $\pm$     9.9  &   -74.4  $\pm$     5.7  & 27&    30.1 $\pm$     9.6 & 27&     14.9 $\pm$      3.5 & 27&   1.52 &   5.14 &   6.37 &  16.50 &  -2.82 &   0 &25,22 &93\\ 
                        GJ 3331 BC &        M3.5V+M4V & 05:06:49 &    -21:35:04 &     33.1 $\pm$      2.7 &    -33.2 $\pm$      2.0 & 30 & 50.7 $\pm$     0.3 & 5 &     23.7 $\pm$      1.7 & 6 &   1.66 &   5.39 &   6.11 &  16.77 &  -2.97 &  20 &2 &96 \\ 
                           GJ 3331 A &              M1V &     05:06:49 &    -21:35:09 &    46.6  $\pm$     0.6 &   -16.3   $\pm$     1.0  &5 &    50.7  $\pm$     0.3 & 5&     21.2 $\pm$      0.9 &6 &   1.42 &   4.32 &   6.12 &  16.73 &  -2.80 &  20 &2 &99 \\ 
                           HIP 25486 &               F7 &     05:27:04 &    -11:54:04 &    17.1 $\pm$     0.03 &   -49.2  $\pm$     0.02 &5 &    37.4  $\pm$      0.3 &5 &    20.2 $\pm$     0.5 & 29&   0.53 &   1.37 &   4.93 &  12.86 &  -3.53 & 162 &28,14,25 &100 \\ 
                           HIP 27321 &              A5V &     05:47:17 &    -51:03:59 &     4.7  $\pm$     0.1  &     83.1 $\pm$     0.2  & 27 &    51.4 $\pm$     0.1 & 27 &     20.0 $\pm$      0.7 & 7&   0.16 &   0.32 &   3.53 &    $-$ &    $-$&   0 & 2,25 &100 \\ 
                          HIP 29964 &             K4V &     06:18:28 &    -72:02:43 &    -7.7  $\pm$     0.1  &    74.4   $\pm$     0.1  & 5 &    25.6  $\pm$     0.2 &5  &     16.2 $\pm$      1.0 &16 &   1.07 &   3.18 &   6.81 &  16.70 &  -2.72 & 400 &28,14,4,25,31& 100 \\ 
                             TWA 22 B &             M6V &     10:17:26 &    -53:54:26 &   -175.8 $\pm$     0.8 &    -21.3 $\pm$     0.8  &24 &     57.0 $\pm$      0.7 & 24&     14.8 $\pm$      2.1 & 24&   1.73 &   6.27 &   7.69 &  19.55 &  -2.89 & 580 & 19,22& 100 \\ 
                             TWA 22 A &             M6V &     10:17:26 &    -53:54:26 &   -175.8 $\pm$     0.8 &    -21.3 $\pm$     0.8  &24 &     57.0 $\pm$      0.7 & 24&     14.8 $\pm$      2.1 & 24&   1.73 &   6.27 &   7.69 &  19.55 &  -2.89 & 580 & 19,22&100\\ 
                       HIP 76629 BC &            M4.5 &     15:38:56 &    -57:42:18 &    -52.9                      &   -106.0                        & 25 &    27.1 $\pm$     0.3 & 5 &      0.1 $\pm$      2.0 & 25 &   1.71 &   5.61 &   9.19 &    $-$ &   -1.6 & 425 &25,2,22 &  100 \\ 
                          HIP 76629 A &              K0V &     15:38:57 &    -57:42:26 &   -49.9  $\pm$     0.06 &   -97.9  $\pm$     0.1  &5 &    27.1 $\pm$     0.3 & 5&      3.1 $\pm$      0.8 &27 &   0.85 &   2.30 &   5.85 &  14.18 &  -3.24 & 280 &28,14,25,31 & 100 \\ 
                           HIP 79881 &               A0 &     16:18:17 &    -28:36:51 &   -31.2 $\pm$     0.3    &  -100.9  $\pm$     0.2  & 27&    24.2 $\pm$     0.2 & 27&    -13.0 $\pm$      0.8 & 7&   0.02 &   0.04 &   4.74 &  $-$ &    $-$ &   0  & 32&    99 \\ 
                         HIP 84586 A &             G5IV &     17:17:25 &    -66:57:03 &   -21.5 $\pm$     0.02 &   -137.3 $\pm$     0.03 &5 &    32.8  $\pm$     0.4 &5 &      5.9 $\pm$      0.2 &27 &   0.81 &   2.17 &   4.70 &  12.78 &  -3.20 & 250 &25 & 100\\ 
                          HIP 84586 B &             K0IV &     17:17:25 &    -66:57:03 &   -21.5 $\pm$     0.02 &   -137.3 $\pm$     0.03 & 5&    32.8  $\pm$     0.4 & 5&      5.9 $\pm$      0.2 & 27&   0.81 &   2.17 &   4.70 &  12.78 &  -3.20 & 250 &25 &  100 \\ 
                        HIP 84586 C &             M3V &     17:17:31 &    -66:57:05 &    -11.0 $\pm$      2.0 &   -143.0 $\pm$      2.0 & 30&    32.8  $\pm$     0.4 & 5&      2.7 $\pm$      1.8 &25 &   1.54 &   5.19 &   7.63 &    $-$ &  -1.45 &  20 &2,25 &100 \\ 
                          HIP 86598 &              F9V &     17:41:48 &    -50:43:28 &     -3.7 $\pm$     1.1   &    -65.7 $\pm$     0.9  & 27&     13.8 $\pm$     0.9 &27 &      1.7 $\pm$      1.7 & 23&   0.55 &   1.37 &   6.99 &  $-$ &  -3.64 & 130 &9 &  85\\ 
                       HIP 88399 A &              F5V &     18:03:03 &    -51:38:54 &     2.3 $\pm$     0.04 &   -86.1  $\pm$     0.03 & 5&    19.8   $\pm$     0.3 &5 &     -0.4 $\pm$      0.5 & 27&   0.43 &   1.10 &   5.91 &    $-$ &  -4.53 & 107&25 &100 \\ 
                         HIP 88399 B &             M2V  &     18:03:04 &    -51:38:56 &     2.3 $\pm$     0.04 &   -86.1  $\pm$     0.03 & 5&    19.8  $\pm$     0.3 & 5&     -2.4 $\pm$      1.3 & 25&   1.52 &   4.23 &   8.27 &    $-$ &  -2.93 &  70&25 & 100 \\ 
                          HIP 88726 A &              A5V &     18:06:49 &    -43:25:30 &    10.7 $\pm$     1.1  &  -106.6  $\pm$     0.5  & 30&     23.9 $\pm$     0.7 & 27&     -7.8 $\pm$      0.4 & 7&   0.22 &   0.55 &   4.39 &    $-$&    $-$ &   0 &11 & 100\\ 
                         HIP 88726 B &              A5V &     18:06:49 &    -43:25:29 &    10.7 $\pm$     1.1  &  -106.6  $\pm$     0.5  & 30&     23.9 $\pm$     0.7 & 27&     -7.8 $\pm$      0.4 & 7&   0.24 &   0.55 &   4.39 &    $-$ &    $-$ &   0 &11& 100 \\ 
                          HIP 89829 &              G5V &     18:19:52 &    -29:16:32 &     4.6   $\pm$     0.1  &   -46.4  $\pm$     0.1  & 5&    12.6  $\pm$     0.3 & 5&     -7.0 $\pm$      2.6 & 25&   0.64 &   1.75 &   7.05 &    $-$ &  -3.23 & 290 & 2,25 & 88 \\ 
                          HIP 92024 A &               A7 &     18:45:26 &    -64:52:16 &     32.4  $\pm$     0.2  &  -149.5  $\pm$     0.2  & 27&    35.0  $\pm$     0.2 & 27&      2.0 $\pm$      4.2 &27 &   0.20 &   0.47 &   4.30 &    $-$ &  -5.70 &   0&25 &  99 \\ 
                       HIP 92024 BC &              K7V &     18:45:36 &    -64:51:45 &  25.9  $\pm$      8.0 &   -184.2 $\pm$      8.0 & 30&    35.0  $\pm$     0.2 & 27&      1.0 $\pm$      3.0 & 25&   1.12 &   3.30 &   6.10 &    $-$ &  -2.98 & 477 &14,25  & 98\\ 
                            HIP 92680 &             G9IV &     18:53:05 &    -50:10:47 &    17.6   $\pm$     1.1  &   -83.6  $\pm$     0.8  & 27&    19.4  $\pm$     1.0  &27 &     -4.2 $\pm$      0.2 &7 &   0.81 &   2.04 &   6.37 &  14.48 &  -3.23 & 279 &14,25&100 \\ 
                           HIP 95270 &             F5.5 &     19:22:58 &    -54:32:15 &    24.5    $\pm$     0.04 &   -82.2  $\pm$     0.03 &5 &    20.6  $\pm$     0.5 & 5&      0.1 $\pm$      0.4 &7 &   0.46 &   1.13 &   5.91 &   $-$ &    $-$ & 117 &28,14,25 & 100 \\ 
                          HIP 99273 &              F5V &     20:09:05 &    -26:13:26 &    40.4   $\pm$     0.04 &   -67.5  $\pm$     0.03 &5 &    19.6  $\pm$      0.3 &5 &     -6.4 $\pm$      1.7 &3 &   0.44 &   1.10 &   6.08 &  12.07 &  -4.90 &  95 & 2& 100 \\ 
                       HIP 102141 B &             M4V &     20:41:50 &    -32:26:10 &    286.2 $\pm$      8.0 &   -377.2 $\pm$      8.0 & 30&     93.5 $\pm$     3.7 & 27&     -5.2  & 25&   1.6  &   5.39 &   4.94 &  15.96 &  -2.63 &   0 &25& 100 \\ 
                        HIP 102141 A &             M4V &     20:41:51 &    -32:26:07 &   270.5 $\pm$     4.6  &   -365.6 $\pm$      3.5 & 27&     93.5 $\pm$     3.7 & 27&     -3.7 $\pm$      3.0 & 15&   1.55 &   5.39 &   4.94 &  15.96 &  -2.63 &   0&25 &   100 \\ 
                          HIP 102409 &             M1V &     20:45:09 &    -31:20:27 &   281.4  $\pm$     0.1  &  -360.1  $\pm$     0.04 & 5&   102.1  $\pm$    0.4 & 5&     -4.5 $\pm$     0.3 & 1 &   1.45 &   4.23 &   4.53 &  15.61 &  -2.77 &  68 & 28,14,25&   100 \\ 
                       HIP 103311 AB &              F8V &     20:55:47 &    -17:06:51 &    58.8 $\pm$     0.8  &   -62.8  $\pm$     0.7 & 27&     21.9 $\pm$     0.8  & 27&     -4.5 $\pm$      2.1 &25 &   0.52 &   1.51 &   5.81 &   $-$ &  -3.41 & 110 &28,25,8&  100 \\ 
                         HIP 112312 A &            M4IV &     22:44:57 &    -33:15:02 &   179.9 $\pm$     0.2   &   -123.3 $\pm$     0.1  & 5&    48.2 $\pm$     0.6 & 5&      3.2 $\pm$      0.5 &20 &   1.52 &   5.14 &   6.93 &  18.26 &  -2.36 &   0 &21,25 & 100 \\ 
                        HIP 112312 B &            M5IV &     22:45:00 &    -33:15:26 &    171.1 $\pm$      1.3 &   -125.2 $\pm$      4.3 & 30 &    48.2  $\pm$     0.6 & 5&     2.0 $\pm$     5.2 &10 &   1.60 &   5.56 &   7.79 &  19.77 &  -2.22 & 336 &14,4,25,26 &  100\\ 
\hline
  \multicolumn{19}{c}{Newly Confirmed Members} \\ \hline
       2MASS J00172353-6645124 &        M2.5 &   00:17:24   &    -66:45:13 &   102.9  $\pm$    1.0 &   -15.0 $\pm$    1.0 &       30 &    25.6 $\pm$    1.7 &       17 &    10.8 $\pm$    0.2 &        12 &   1.54 &   4.65 & 7.70& 19.12 &  -3.00 & $-$&  $-$  & 100 \\ 
                      GJ 2006 A           &               M4   &  00:27:50   &    -32:33:06  &    99.2  $\pm$    1.3 &   -61.3 $\pm$    2.6 &       30 &    30.1 $\pm$    2.5 &     17 &     9.5 $\pm$    0.3 &       11 &   1.38 &   4.94 & 8.01 &  19.48 &  -2.18 & $-$ & $-$ &100 \\ 
                      GJ 2006 B          &             M3.5 &  00:27:50   &    -32:33:24  &    117.2 $\pm$    4.1 &   -31.5 $\pm$    5.8 &    30 &    31.8  $\pm$    2.5 &     17 &     8.5 $\pm$    0.2 &       12 &   1.41 &   5.13 & 8.12 & 18.87 &  -2.20 & $-$ &  $-$ &100\\ 
       2MASS J16572029-5343316 &         M3  & 16:57:20   &    -53:43:32  &   -13.0  $\pm$    6.3 &   -85.1 $\pm$    2.2 &      30 &    19.4 $\pm$    0.7 &     5 &      1.4 $\pm$     0.2 &   12 &   1.46 &   4.62 & 7.79 & $-$ &  -3.23 & $-$ & $-$ & 100 \\ 
                      CD-54 7336       &              K1V  &  17:29:55   &    -54:15:49  &    -9.8   $\pm$    3.2 &   -60.0 $\pm$    1.7 &      30 &    14.4 $\pm$    0.2 &  5 &     -0.2 $\pm$     0.9 &   3 &   0.77 &   2.25 & 7.36 &   $-$ &  -3.13 & 360 & 2,25 & 97  \\ 
                   CD-31 16041       &             K8V  & 18:50:44   &    -31:47:47  &    16.4  $\pm$    1.6 &   -72.8 $\pm$    1.1 &       30 &    20.1 $\pm$    0.3 &    5 &     -6.0 $\pm$     1.0 &     25 &   1.06 &   3.73 &  7.46 & 18.04 &  -2.79 & 492 & 2,25 & 100 \\ 
                TYC 7443-1102-1    &            K9IV  & 19:56:04   &    -32:07:38  &    31.9  $\pm$    1.4 &   -65.1 $\pm$    1.2 &     30 &    19.9   $\pm$    0.3 &  5  &    -7.1 $\pm$    2.2 &   10 &   1.36 &   3.74 & 7.85 &  19.09 &  -2.89 & 110 & 13,9  & 99 \\ 
               UCAC3 124-580676 &               M3  & 20:10:00   &    -28:01:41  &    40.7   $\pm$    3.0 &   -62.0 $\pm$    1.7 &    30 &    20.9   $\pm$    1.3 &    17 &     -5.8 $\pm$     0.6 &     12 &   1.50 &   5.26 & 7.73 & 18.43 &  -2.66 & $-$ & $-$ & 99\\ 
        2MASS J20333759-2556521 &     M4.5  & 20:33:38   &    -25:56:52  &    52.8   $\pm$    1.7 &   -75.9 $\pm$    1.3 &   30 &    20.7   $\pm$    1.4 &   17  &     -6.0 $\pm$     0.5 &  12 &   1.71 &   5.99 & 8.88 & 20.25 &  -3.04 & $-$ &  $-$ & 100\\ 
       2MASS J21212873-6655063 &      K7V  &  21:21:29  &    -66:55:06  &    97.2   $\pm$    1.1 &  -104.1 $\pm$    1.6 &  30 &    31.1   $\pm$    0.8 &   5 &      3.3                         &     25 &   1.34 &   3.59 & 7.01 &  $-$ &  -2.92 &  15 & 25 &100\\ 
                     CPD-72 2713 &                K7V   &  22:42:49   &    -71:42:21  &    92.7   $\pm$    0.8 &   -51.1  $\pm$    0.8 &  30 &    27.4   $\pm$    0.3 &   5 &      8.6 $\pm$     0.5 &   25 &   1.32 &   3.67 & 6.89 & 17.55 &  -2.80 & 440 & 2,25 &100  \\ 
                     BD-13 6424 &                  M0V  & 23:32:31   &    -12:15:51  &   137.4  $\pm$    1.0 &   -81.0  $\pm$    1.0 &  30 &    36.0    $\pm$    0.5 &  5 &      1.8 $\pm$     0.7 &    25 &   1.74 &   4.07 & 6.57 & 17.82 &  -3.68 & 185 & 2,25  & 100\\ 
 \hline 
  \multicolumn{19}{c}{Probable Members} \\ \hline
  \multicolumn{19}{c}{Group1: young, but missing $\pi$ or RV} \\
                 GSC 08350-01924 &             M3V  & 17:29:21 &   -50:14:53    &    -5.8 $\pm$    1.5 &   -62.7 $\pm$    5.1 &   30    &    $-$ &               $-$ &     0.3 $\pm$    1.1 &       12 &   1.46 &   4.87 & 7.99 &  $-$  &  -2.94 &  50 & 2  &88 \\ 
                      HD 161460      &             K0IV & 17:48:34   &   -53:06:43    &    -3.6 $\pm$    1.0 &   -58.4 $\pm$    1.3 &     30  &  $-$ &    $-$             &    -0.2 $\pm$    1.5 &    23  &  0.97 &  2.31 & 6.78 &   $-$ &  -3.14 & 320 & 2,25 & 89 \\ 
                    Smethells 20      &             M1V  & 18:46:53 &   -62:10:37     &    13.6 $\pm$    1.4 &   -79.4 $\pm$    1.4 &    30  &      $-$  &           $-$  &     0.3 $\pm$    3.2 &                      10 &   1.24 &   3.98 & 7.85 &     $-$ &  -3.00 & 332 & 2,25 &97 \\ 
                           AZ Cap       &               K7  & 20:56:03   &   -17:10:54     &    57.6 $\pm$    1.1 &   -59.9 $\pm$    1.2 &     30 &   $-$ &              $-$ &    -6.9                      &        25 &   1.12 &   3.44 & 7.04 &    $-$ &  -3.25 & 235.5 & 25,14 & 99 \\ 

\multicolumn{19}{c}{Group2: probably young, but missing $\pi$ or RV} \\    
                         2MASS J00281434-3227556            &               M5 & 00:28:14   & -32:27:56   &   110.1 $\pm$    1.8 &   -43.0 $\pm$    3.3 &      30 &      $-$  &                       $-$ &     5.9 $\pm$    3.4 &        12 &   1.58 &   5.95 & 9.28 & 20.85 &  -2.55 & $-$ & $-$ & 89 \\ 
                 PYC J00390+1330 &  M4         & 00:39:03   & +13:30:17   &    85.5 $\pm$    3.2 &   -68.0 $\pm$    3.9 &      30 &     $-$ &                       $-$ &      $-$&                       $-$ &   1.60 &   5.64 & 10 06 &  21.29 &  -2.74 & $-$ & $-$ & 97\\ 
                     BD+17 232A       &             $-$  & 01:37:39   & +18:35:33  &    68.6 $\pm$    0.8 &   -47.3 $\pm$    0.6 &        30 &     $-$ &     $-$&     3.2 $\pm$    1.0 &   18 &   1.03 &   3.86 & 6.72 &  14.37 &  -2.62 & $-$ & $-$ & 97 \\ 
                 UCAC3 176-23654 &               M3 & 05:34:00   & -02:21:32  &    12.3 $\pm$    1.2 &   -61.3 $\pm$    2.4 &       30 &      $-$ &                     $-$ &    21.0 $\pm$    0.2 &        12 &   1.49 &   4.72 & 7.70 &    $-$ &  -2.57 &   0 & 32 &96 \\ 
        2MASS J08173943-8243298 &    M3.5 & 08:17:39   & -82:43:30  &   -80.3 $\pm$    1.1 &   102.5 $\pm$    0.8 &      30 &      $-$ &                       $-$ &    17.5 $\pm$    1.6 &        12 &   1.58 &   5.03 & 6.59 & 17.59 &  -2.94 &   0 & 32 & 94 \\ 
                UPM J1354-7121        &      M2.5 & 13:54:54   & -71:21:48  &  -165.0 $\pm$    8.0 &  -132.7 $\pm$    8.0 &     30 &      $-$  &                  $-$ &     5.7 $\pm$    0.2 &        12 &   1.48 &   4.57 & 7.67 & 18.53 &  -3.10 & $-$ & $-$ & 100 \\ 
        2MASS J17150219-3333398  &    M0   & 17:15:02   & -33:33:40   &     7.8   $\pm$     1.0 & -175.9 $\pm$  1.2  &       30 &      $-$  &                  $-$ &    -14.6 $\pm$ 3.5 &          12 &  1.41 &    3.87 &    7.07  & $-$    & -2.99 & $-$ & $-$ & 87 \\
        2MASS J18420694-5554254 &   M3.5 & 18:42:07   & -55:54:26   &     9.7 $\pm$    12.1 &   -81.2 $\pm$    2.8 &        30 &     $-$ &                 $-$ &      0.3 $\pm$     0.5 &        12  &   1.58 &   4.95 & 8.58 &  19.70 &  -2.72 &   0 & 32  & 98 \\ 
         2MASS J19102820-2319486 &      M4 & 19:10:28   & -23:19:49   &    16.6 $\pm$    1.4 &   -51.8 $\pm$    1.4 &      30 &     $-$ &         $-$ &     -8.0 $\pm$     0.8 & 12 &   1.53 &   5.01 & 8.21 & 19.08 &  -2.69 & $-$ &  $-$ & 86  \\ 
         2MASS J19243494-3442392 &      M4 & 19:24:35   & -34:42:39   &    22.1 $\pm$    1.8 &   -71.7 $\pm$    1.8 &       30 &     $-$ &      $-$ &     -3.2 $\pm$     0.3 &  12 &   1.58 &   5.52 & 8.79 &  20.00 &  -3.11 & $-$ & $-$ & 94 \\ 
    UCAC3 116-474938                   &     M4  & 19:56:03  & -32:07:19   &   35.2 $\pm$     1.8 &   -59.9 $\pm$    1.5 &     30 &        $-$ & $-$    & -2.8 $\pm$ 1.8        & 12   & 1.56  &    5.12 &  8.1   & 19.60 & -2.73 & 0 &  9 & 81 \\
                 GSC 06354-00357         &      M2 & 21:10:05   & -19:19:57   &    89.0 $\pm$    0.9 &   -89.9 $\pm$    1.8 &      30 &   $-$&               $-$ &    -5.5 $\pm$    0.5 &  12 &   1.52 &   4.46 & 7.20 & 18.58 &  -2.83 & $-$ & $-$ & 100  \\ 
\multicolumn{19}{c}{Group3: young with full 6 kinematic parameters, but low membership probability} \\ 
                TYC 6872-1011-1 &             M0V  & 18:58:04 & -29:53:05 &    12.2 $\pm$    1.3 &   -45.7 $\pm$    2.5 &        30 &    12.8 $\pm$    0.4 &   5 &     -4.9 $\pm$     1.0 &        25 &   2.16 &   3.78 & 8.02 &   $-$ &    $-$ & 483 & 2,25 & 60\\ 
 \hline 
    \end{tabular} 
    \begin{tablenotes}
    \item[a] If multiple data are available, weighted mean is used.
    \item[b] The membership probability for BPMG is calculated using the improved models (the new field star model and uniformly distributed BPMG model in XYZ using \listexc--\casev).
    
       \item Notes.
   \item References to the table:
(1) \citet{chu11}; (2) \citet{sil09}; (3) \citet{des15}; (4) Fern\`{a}ndez, Figueras \& Torra (2008); (5) \citet{gai16}; (6) Gizis, Reid \& Hawley (2002); (7) \citet{gon06};
   (8) \citet{kai04}; (9) \citet{kis11}; (10) \citet{kor13}; (11) \citet{kra14}; (12) \citet{mal14}; (13) \citet{mcc12};
   (14) \citet{men08}; (15) \citet{mon01a}; (16) \citet{mon01b}; (17) \citet{rie14}; (18) \citet{sch10}; (19) \citet{shk11};
   (20) \citet{shk12}; (21) Song, Bessell \& Zuckerman (2002); (22) \citet{son03}; (23) \citet{son12}; (24) \citet{tei09}; (25) \citet{tor06};
   (26) van Altena, Lee \& Hoffleit (1995); (27) \citet{lee07}; (28) \citet{wei10}; (29) White, Gabor \& Hillenbrand (2007); (30) \citet{zac12}; (31) \citet{zuc01a}; (32) Internel data

    \end{tablenotes}
 \end{threeparttable} 
  \end{table*}

\section{Summary and conclusions}

Deploying the same formulation that BANYAN II used, we examine the impact of three  assumptions on models in the MG membership probability calculation: accepted initial member lists, distribution models of MG members, and distribution models of field stars.
Reassessment of membership of BPMG members in the \listone\ results in a refined kinematic model for BPMG.
Depending on the membership assessment criteria (exclusive and inclusive), membership probabilities of stars in a test set (the \listthree) change up to $\sim$40 per cent.
Lacking evidence of a central concentration of BPMG members in $XYZ$, we suggest to use the uniform distribution model in $XYZ$.
This uniform spatial distribution model changes the membership probabilities of the test stars up to $\sim$80  per cent compared to the Gaussian distribution.
For field star models, assuming the uniform distribution in $XYZ$ is more realistic compared to the Gaussian distribution model;
the uniform distribution model expects more field stars at larger distance, while the Gaussian model expects the maximum stellar number density at $\sim$120 pc, which seems to be artificial.
In $UVW$, field stars show distinct subgroups, and the model properties of these subgroups are obtained using a Gaussian mixture model.
Combined effect of these model modifications show changes in membership probabilities of the test stars up to $\sim$80 per cent.
These comparisons show significant membership probability changes especially for some marginal members, indicating the sensitivity to prior knowledge on the MG membership calculation and the importance of using reliable models.

We confirm 57 (51, if we exclude 6 classical members showing ambiguity in youth) BPMG members from the \listthree.
Only about 90 stars from the \listthree\ seem to be kinematically associated with BPMG (p $>$80 per cent), and 51 (12 stars are new compared to the \listone) out of these $\sim$90 stars show unambiguous signs of youth with 6 full kinematic parameters, which allow us to confirm them as bona fide BPMG members.
Additionally, we suggest 17 probable BPMG members.

In this study, we considered only kinematic properties in the MG membership probability calculation.
Because the number density of field stars is much larger than those of MG members and there are many field stars with similar $UVW$ to that of MGs, the contamination (old interlopers) rate has to be significantly large without considering age-related information.
In the future, we will formally incorporate the age-related information into the Bayesian scheme developed in this study to provide a more reliable MG membership calculation (Lee \& Song in preparation).

\clearpage

\section*{Acknowledgements}

We thank the anonymous referee for valuable comments and suggestions that helped to significantly increase the quality of this work.












\bsp	
\label{lastpage}
\end{document}